\documentclass[fleqn,usenatbib,useAMS]{mnras}

\usepackage[dvipsnames]{xcolor}
\usepackage{tikz,hyperref}

\definecolor{lime}{HTML}{A6CE39}
\DeclareRobustCommand{\orcidicon}{
	\begin{tikzpicture}
	\draw[lime, fill=lime] (0,0) 
	circle [radius=0.13] 
	node[white] {{\fontfamily{qag}\selectfont \tiny ID}};
	\draw[white, fill=white] (-0.0625,0.095) 
	circle [radius=0.007];
	\end{tikzpicture}
	\hspace{-2mm}
}

\foreach \x in {A, ..., Z}{\expandafter\xdef\csname orcid\x\endcsname{\noexpand\href{https://orcid.org/\csname orcidauthor\x\endcsname}
			{\noexpand\orcidicon}}
}
\usepackage{newtxtext,newtxmath}

\usepackage{booktabs}
\usepackage[T1]{fontenc}
\usepackage{ae,aecompl}

\usepackage{graphicx}	% Including figure files
\usepackage{amsmath}	% Advanced maths commands
\usepackage{booktabs}
\usepackage{times}
\usepackage{subcaption} % para subfiguras
\input{epsf}

\title[Hot bubbles in PNe (III)]{Formation and X-ray emission from hot bubbles in planetary nebulae - III. The impact of [Wolf-Rayet]-type winds}

\author[R.~Orozco-Duarte et al.]{Rogelio~Orozco-Duarte\thanks{E-mail: r.orozco@irya.unam.mx}$^{1}\orcidA$, Jes\'{u}s~A.~Toal\'{a}$^{1}\orcidB$, S. Jane Arthur$^{1}\orcidD$, Janis~B.~Rodr\'{i}guez-Gonz\'{a}lez$^{2}\orcidC$,  
\newauthor Luke~Conmy$^{3}$ and Rolf~Kuiper$^{3}\orcidE$\\
$^{1}$Instituto de Radioastronom\'{i}a y Astrof\'{i}sica, Universidad Nacional Aut\'{o}noma de M\'{e}xico, 58089 Morelia, Michoac\'{a}n, Mexico\\
$^{2}$Instituto de Astrof\'{i}sica de Andaluc\'{i}a, CSIC, Glorieta de la Astronom\'{i}a S/N, Granada E-18008, Spain\\
$^{3}$Fakultät für Physik, Universität Duisburg-Essen, Lotharstra$\beta$e 1, D-47057 Duisburg, Germany.
}

\date{\today}%Accepted XXX. Received YYY; in original form ZZZ}

\pubyear{2026}

\begin{document}
\label{firstpage}
\pagerange{\pageref{firstpage}--\pageref{lastpage}}
\maketitle

\begin{abstract}
We use radiation-hydrodynamical simulations to investigate the formation and synthetic X-ray emission of hot bubbles within planetary nebulae (PNe) driven by the powerful winds of H-deficient, [Wolf-Rayet]([WR])-type stars. Our models, based on {\sc mesa} stellar evolution tracks for 1--3 M$_{\odot}$ progenitors, adopt a recent mass-loss rate prescription for [WR] stars and incorporate the enhanced radiative cooling of their C-rich material, comparing the results against standard H-rich PN models. The enhanced mass-loss in the [WR] models leads to an accelerated post-AGB evolution and a subsequent delay in hot bubble formation compared to their H-rich counterparts, as suggested by a previous work. 
By computing synthetic X-ray spectra that account for the mixed H-rich and H-deficient gas phases, we find that models incorporating [WR] winds exhibit significantly higher X-ray luminosities ($L_\mathrm{X}$) than their H-rich counterparts, but the emissivity-weighted plasma temperature of the X-ray-emitting gas converge to values of $T_\mathrm{X} = [1-3] \times 10^{6}$~K, regardless of whether the system follows a [WR]-type or an H-rich post-AGB evolutionary path. Our results reinforce previous suggestions that mixing is a key mechanism in generating the observed soft X-ray emission even for PN hosting [WR] central stars.
\end{abstract}


%%%%%%%%%%%%%%%%% BODY OF PAPER %%%%%%%%%%%%%%%%%%

\section{Introduction}\label{introduction}
\label{sec:intro}

Planetary nebulae (PNe) represent a late evolutionary stage of low- and intermediate-mass stars ($M_\mathrm{ZAMS} \sim1$--8 M$_{\odot}$)\footnote{ZAMS stands for zero age main sequence.}, marking the transition from the asymptotic giant branch (AGB) into the white dwarf phase. After ejecting its outer layers, the star enters the post-AGB phase developing a fast stellar wind \citep[$\varv_\infty \gtrsim 1000$ km~s$^{-1}$;][]{Kwok1978,Guerrero2013} that interacts with the previously ejected, dense and slow-moving AGB material 
\citep[$\varv_\mathrm{AGB} \approx 10$ km~s$^{-1}$, $\dot{M}_\mathrm{AGB} \lesssim 10^{-5}$~M$_\odot$ yr$^{-1}$; e.g.,][and references therein]{Hofner2018,Ramstedt2020}, 
forming a characteristic shell of ionised gas \citep[e.g.,][]{Kwok1978,Kwitter2022}. 
One of the direct consequences of this wind-wind interaction is the formation of hot, X-ray emitting bubbles within the nebular interior, a phenomenon observed in several well-studied PNe \citep[e.g.,][]{Kastner2000,Chu2001,Guerrero2000,Gruendl2006,Montez2005}.

The hot bubbles arise because of the interaction of the highly supersonic, tenuous post-AGB wind with the dense material ejected earlier in the AGB phase. A characteristic two-shock structure is set up, where the outer shock sweeps up the AGB material, while an inward-facing shock thermalises the fast wind. The plasma temperature in the hot, shocked wind bubble can be calculated using the assumption that the inner shock is adiabatic, $T = 2.26\times10^7 \mu \varv_{\infty,3}^2$, where $\mu$ is the mean mass per particle in the shocked wind and $\varv_{\infty,3} = \varv_\infty/(1000\,\mbox{km\,s$^{-1}$})$ is the stellar wind velocity \citep[][]{Dyson1997}. Nevertheless, observations of PNe report the presence of hot bubbles emitting soft X-ray emission (0.3--2.0 keV) with characteristic temperatures of only $T_\mathrm{X} \approx [1-3] \times 10^{6}$~K \citep[e.g.,][and references therein]{Ruiz2013}.

Thermal conductivity between the hot interior and the cooler ($10^{4}$~K) surrounding photoionised nebula has been proposed as a key mechanism for modifying the temperature and density structure of the hot bubbles \citep[e.g.,][]{Weaver1977,Soker1994}. 
Thermal conduction leads to an evaporative flow from the nebular shell into the hot bubble, forming an interface region where the gas has intermediate temperatures and enhanced densities, which dominates the observed X-ray emission \citep[e.g.,][]{Arthur2012}. The impact of thermal conduction on the X-ray properties of hot bubbles in PNe was investigated using 1D numerical simulations by \citet{Steffen2008}, incorporating a detailed evolution of stellar wind parameters, which resulted in good agreement with observations. However, the strongest argument against thermal conductivity is that, if present, a magnetic field may suppress thermal conduction by electrons \citep{Dyson1981}, which would result in higher than expected temperatures and lower soft X-ray fluxes.

Multi-dimensional numerical simulations by \citet{ToalaArthur2014,ToalaArthur2016} demonstrated that hydrodynamical mixing, driven by instabilities in the wind-wind interaction region, can produce effects similar to those of thermal conductivity \citep[see also][]{Stute2006,ToalaArthur2018}. Those simulations also reproduced the observed properties of hot bubbles in PNe, suggesting that hydrodynamical instabilities may play a crucial role in shaping their X-ray emission when thermal conduction is suppressed or inefficient.

The standard treatment of thermal conduction in the astrophysical literature assumes a fully ionized pure hydrogen plasma and uses the description of \citet{Spitzer1962} to evaluate the diffusion coefficient for charge $Z=1$ \citep[e.g.,][]{Weaver1977,Steffen2008,ToalaArthur2011}. However, one of the key findings of the Chandra Planetary Nebula Survey \cite[{\sc ChanPlaNS};][]{Kastner2012,Freeman2014,Montez2015} is that a significant fraction of X-ray-emitting PNe host hot bubbles associated with H-deficient central stars (CSPNe), those denoted as [Wolf-Rayet]-type\footnote{Square brackets are used to differentiate between their massive counterparts. This definition was originally proposed by \citet{vaderHucht1981}.}.

A widely accepted channel to produce [WR]\footnote{Hereinafter [WR] will stand for [Wolf-Rayet].}  CSPN is the born-again scenario, in which a late thermal pulse occurs during the post-AGB evolution leading to the ingestion and burning of residual hydrogen and an H-deficient surface composition \citep{Iben1983,MillerBertolami2024}. This evolutionary path contrasts with the standard H-rich post-AGB evolution and naturally motivates exploring how an H-deficient wind affects the formation and X-ray emission of hot bubbles. \citet{Sandin2016} extended the heat conduction formulation to take into account the chemistry dependence and applied it to the case of PNe that show H-deficient, chemically stratified hot bubbles. The H-deficient case has a lower diffusion coefficient than the H-rich case, which results in a thinner, and therefore higher temperature, interface region. Additionally, the enhanced radiative cooling in gas rich in C and O delays the formation of a hot bubble \citep{Mellema2002,Sandin2016}.

The study of diffuse X-ray emission produced by the winds of H-deficient [Wolf-Rayet]-type stars was subsequently advanced by \citet{Heller2018}, who made a very detailed comparison between synthetic and observed high-dispersion X-ray spectra of the hot bubble of the [WR] PN BD+30$^\circ$3639. They used an analytical, self-similar description for the temperature and density profile of the hot bubble, which includes the thermal conduction prescription of \citet{Zhekov1996}, together with abundances obtained from photosphere and wind analyses of BD+30$^\circ$3639, to construct the ionisation state and X-ray spectra of a selection of the most relevant chemical species. An extension of that work has been recently presented by \citet{Schonberner2024}, where they used a reference 0.595 M$_\odot$ post-AGB stellar evolution model to make predictions of hot bubbles produced by winds from [WR]-type stars by varying the stellar wind luminosity, chemical composition, and the evolutionary speed across the Hertzsprung-Russell (HR) diagram. In that work, the stellar wind velocity and mass-loss rates of the 0.595 M$_\odot$ post-AGB model are enhanced by factors of 10 and 100 to imitate the large mechanical wind luminosity of [WR] CSPNe. \citet{Schonberner2024} found that the hot bubble formation and evolution of [WR] PNe is impacted by the high radiative cooling due to metals such as C and O in their winds \citep[cf.][]{Mellema2002}. Consequently, the formation of their hot bubbles is delayed compared with models including H-rich CSPN.

Currently, there are no specific stellar evolution models that describe the formation of PNe around H-deficient [WR] stars and most of what is known about the evolution of low-mass stars is through H-rich stellar evolution models \citep[e.g.,][]{MB2016}. The main problem is characterising the mass-loss rate, which, along with the initial mass of the star, dictates the final destiny of the evolving star. In an attempt to discern the properties of the winds of WR-type stars, \citet{Toala2024} presented a joint analysis of low-mass and massive stars. Taking advantage of the improvement in distance estimations through {\it Gaia} data \citep{BJ2021} in combination with detailed non-LTE modelling through the {\sc PoWR} code\footnote{\url{https://www.astro.physik.uni-potsdam.de/PoWR/}} \citep{Grafener2002,Hamann2004}, these authors parametrised the mass-loss rate of low-mass and massive WR-type stars. \citet{Toala2024} found that WR-type stars define what they call a fundamental plane established by the luminosity, effective temperature and the mass-loss rate ($L$, $T_\mathrm{eff}$, $\dot{M}$).

In this paper, we use radiation-hydrodynamical simulations to explore the formation and synthetic X-ray emission of hot bubbles in PNe created by the impact of the powerful winds from [WR] stars. We include a detailed evolution of the stellar wind parameters for H-deficient stars following the mass-loss rate predictions from \citet{Toala2024}. Comparisons with the evolution of standard H-rich CSPN allow us to assess the energetics imprinted by [WR] winds, the formation of hot bubbles in PNe, and their predicted X-ray emission.

This paper is organized as follows. In Section~\ref{sec:methods} we describe the stellar evolution models including the approximations for [WR] winds. The results from radiation-hydrodynamical simulations, which include the description of the evolution of hot bubbles and their synthetic X-ray emission, are presented in Section~\ref{sec:results}. A discussion is presented in Section~\ref{sec:discusion} and our conclusions are summarised in Section~\ref{sec:summary}.

\section{Methodology}
\label{sec:methods}

We follow a similar methodology to that presented in previous publications in this series \citep[see][]{ToalaArthur2011,ToalaArthur2014,ToalaArthur2016,ToalaArthur2018}. Stellar evolution models are used to calculate the detailed evolution of the stellar wind parameters and ionising photon rate of low-mass stars evolving from the AGB into the post-AGB phase. Hydrodynamical simulations of the circumstellar medium formed as a result of mass loss during the AGB phase are calculated in 1D and are subsequently remapped into 2D grids to continue the post-AGB evolution. Finally, the synthetic X-ray emission is calculated following a post-processing methodology. % described in Section~\ref{sec:computing_synthetic}.

\subsection{Radiation-hydrodynamical calculations}
\label{sec:simulations}

The numerical simulations presented here are performed with the extensively tested {\sc pluto} code \citep{Mignone2007}\footnote{\url{https://plutocode.ph.unito.it/}}. We run simulations adopting a finite-volume formulation with a Godunov scheme employing an HLLC approximate Riemann solver \citep{Toro1994}. We perform 1D and 2D simulations in spherical coordinates.

Our simulations are conducted on uniform grids with different dimensional setups. The 1D simulations are performed on a radial grid extending up to 4\,pc, consisting of 6000 equidistant zones, resulting in a cell resolution of $\Delta r = 6.67 \times \mathrm{10^{-4}}$\,pc. These simulations are used to study the global impact and general properties of the formation and evolution of hot bubbles.

2D numerical simulations are used to make predictions as to the expected X-ray emission from hot bubbles in PNe as a consequence of hydrodynamical instabilities. In these simulations, the computational domain spans a physical size of 1.0\,pc in both the radial ($r$) and angular ($\theta$) directions, maintaining a consistent resolution across both coordinates of $1000 \times 1000$ equidistant zones.

The free-streaming stellar wind is injected into the 10 innermost cells of the computational domain, where the density is determined from the mass-loss rate $\dot{M} (t)$ and stellar wind velocity $\varv(t)$ using:
\begin{equation}\label{eqn:1}
    \rho(r,t) =  \frac{\dot{M}(t)}{4 \pi r^{2} \varv (t)}.
\end{equation}
\noindent Both quantities, $\dot{M}(t)$ and $\varv(t)$, are functions of time that follow from detailed stellar evolution models (see Section~\ref{sec:windparams}).

\begin{figure}
\begin{center}
\includegraphics[width=\linewidth]{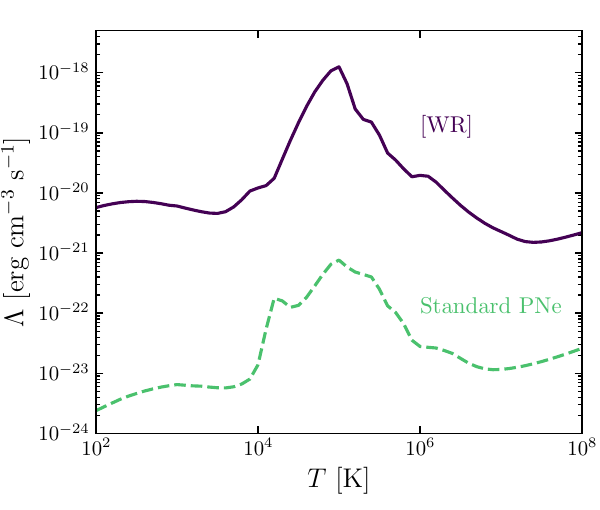}
\end{center}
\caption{Cooling rates as function of gas temperature used in the {\sc pluto} simulations presented in this work. The dashed-line curve was computed adopting standard H-rich PNe abundances while the solid line represents a cooling function computed for H-deficient, C-rich [WR] abundances.}
\label{fig:cooling}
\end{figure}

To model the impact of the ionising flux, we employ the {\sc Sedna} module, a hydrogen photoionisation solver developed by \citet{Kuiper2020} and integrated into the {\sc pluto} code. {\sc Sedna} uses a hybrid approach that combines ray-tracing of direct UV photons from a point source with a flux-limited diffusion treatment of diffuse radiation arising from direct recombinations into the hydrogen ground state. The module solves the ionisation balance using a time-dependent rate equation formalism that includes photoionisation, radiative recombination, and collisional ionisation processes. The evolution of the ionisation fraction directly feeds into the gas thermodynamics by modifying both the mean molecular weight and temperature, thereby enabling a self-consistent treatment of ionisation-driven feedback. {\sc Sedna} also supports both the on-the-spot approximation and a more complete treatment of the diffuse ionising field, making it well-suited for multidimensional applications involving sharp ionisation fronts or shadowed regions.

The hydrodynamic conservation equations solved by {\sc pluto} include radiative cooling and heating source terms. We generate tailored radiative cooling curves, which take into account both the modified abundances and the presence of a central source of ionizing photons, using the spectral synthesis code for astrophysical plasmas {\sc Cloudy}  \citep[version 17;][]{Ferland2017}. {\sc Cloudy} calculates the ionization, thermal, and chemical state of a dilute gas, that may also be exposed to an external radiation source, by simultaneously solving the equations of statistical equilibrium, ionization balance and heating and cooling processes \citep[see][]{Osterbrock2006}. It incorporates an exhaustive atomic database that is subject to continuous revisions to reflect the latest developments in the astrophysical literature. The valid temperature range of the {\sc Cloudy} code is between 10 and $10^9$~K. At temperatures above $10^5$~K collisional processes dominate. The presence of an external radiation source affects the ionisation balance, and therefore the cooling rate, in lower temperature gas. Points on the cooling curves (see Fig.~\ref{fig:cooling}) are generated by summing all of the individual contributions to the cooling from all atomic species and processes in a one-zone, constant density model at a specified kinetic temperature. Free-free and free-bound cooling from heavy elements such as Oxygen and Carbon dominate the cooling of H-deficient gas above $10^6$~K since the amount of cooling per nucleus of these elements is more than an order of magnitude higher than that of Hydrogen or Helium. At temperatures below $10^4$~K, collisionally excited lines of singly ionized metals are important coolants. At lower temperatures, the inclusion of a cosmic ray background makes cooling by C~{\sc ii} an important contributor. In our calculations, we assume an ionisation source with effective temperature of 60\,kK, a mean gas density of 100\,cm$^{-3}$, and an ionising flux $\phi = Q_0/4\pi r^{2}=10^{10}$\,cm$^{-2}$s$^{-1}$, which are typical values for PNe.

Two different sets of abundances are considered (see Table~\ref{tab:abund}): (i) the standard (H-rich) PN abundance set predefined in {\sc Cloudy}, which corresponds to averaged abundances of a sample of 41 PNe listed in \citet{Aller1983} and (ii) the abundance set defined for C-rich [WR] winds listed in table 1 of \citet{Schonberner2024}. The mean atomic weight per particle in the fully-ionised H-rich case is $\mu = 0.62$, while that of the H-deficient abundance set is $\mu = 1.47$. 

The resultant cooling curves are presented in Fig.~\ref{fig:cooling}. Similarly to the collisional ionisation equilibrium cooling curves presented in \citet{Mellema2002}, the cooling produced by a H-deficient material with C-rich [WR] abundances can be about 4 orders of magnitude larger than for typical PNe abundances.

\begin{table}
\centering
\caption{Abundance set (by number) for standard PN and those of [WR] stars$\dagger$.}
\label{tab:abund}
\begin{tabular}{lcc}
\hline
Atom & PN                 & [WR]                     \\
     & \citep{Aller1983}  &  \citep{Schonberner2024} \\
\hline
He & 11.00 & 12.72  \\
C  & 8.89  & 12.30  \\
N  & 8.26  & 7.56   \\
O  & 8.64  & 11.22  \\
F  & 5.48  & 6.10   \\
Ne & 8.04  & 10.55  \\
Na & 6.28  & 7.87   \\
Mg & 6.20  & 9.12   \\
Al & 5.43  & 8.01   \\
Si & 7.00  & 9.01   \\
P  & 5.30  & 6.99   \\
S  & 7.00  & 8.58   \\
Cl & 5.23  & 6.86   \\
Ar & 6.43  & 8.00   \\
K  & 5.08  & 6.66   \\
Ca & 4.08  & 7.90   \\
Fe & 5.70  & 9.04   \\
Ni & 4.26  & 7.79   \\
\hline
\end{tabular}
\begin{flushleft}
$\dagger$The abundances are listed in the $12+\log_{10}\mathrm{H}$ format.
\end{flushleft}
\end{table}

\begin{figure}
    \centering
    \includegraphics[width=1.0\linewidth]{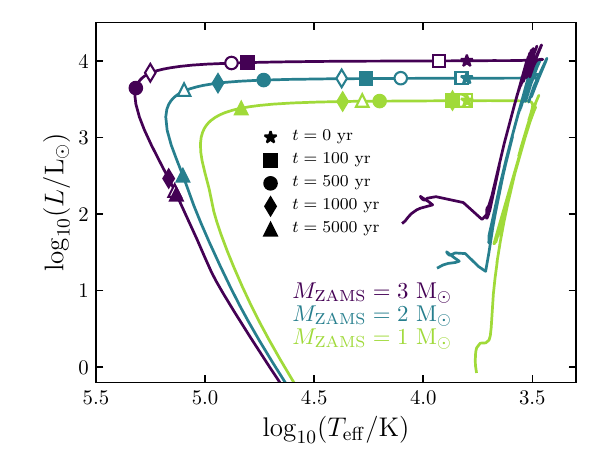}
    \caption{HR diagram of the 1, 2, and 3 M$_\odot$ stellar evolution models created with {\sc mesa}. Symbols on the tracks represent times into the evolution of the post-AGB phase. Filled and open symbols correspond to [WR] and H-rich post-AGB evolution, respectively.}
    \label{fig:HRD}
\end{figure}

\begin{figure*}
    \centering
    \includegraphics[width=0.5\linewidth]{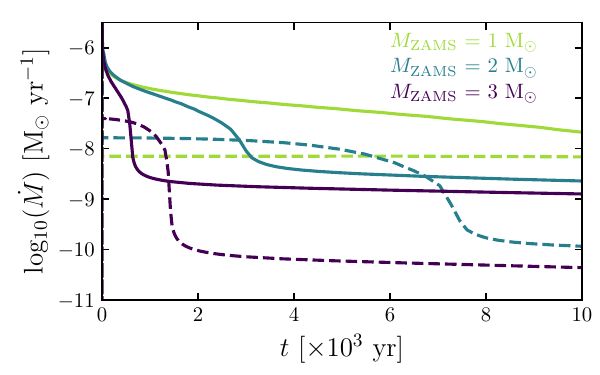}~   
    \includegraphics[width=0.5\linewidth]{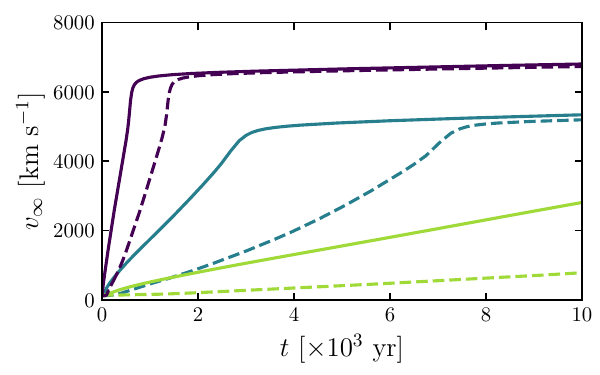}\\
    \includegraphics[width=0.5\linewidth]{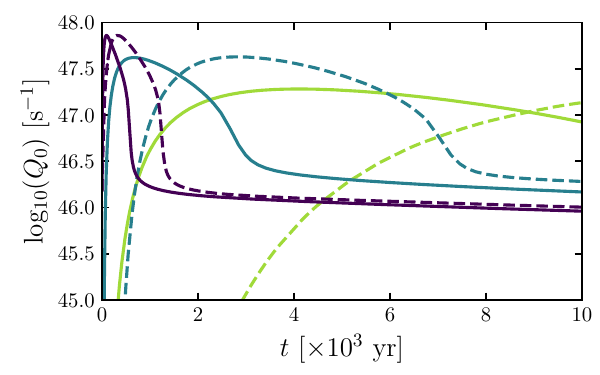}~
    \includegraphics[width=0.5\linewidth]{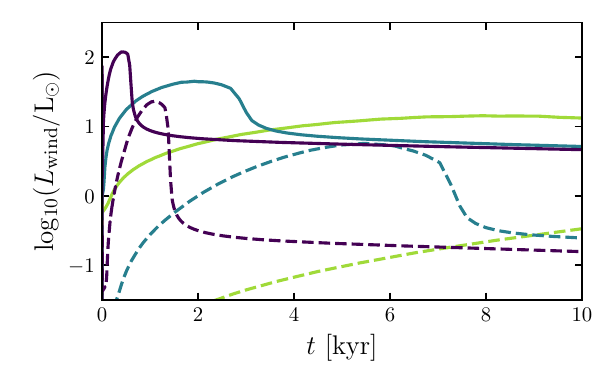}
    \caption{Mass-loss rate ($\dot{M}$, top-left), stellar wind velocity ($\varv_\infty$, top-right), ionising photon rate ($Q_0$, bottom-left), and wind mechanical luminosity ($L_\mathrm{wind}$, bottom-right) during the first 10 kyr of post-AGB evolution. Solid and dashed lines represent [WR] and H-rich models, respectively. Different colours represent different initial masses as indicated in the top-left panel.}
    \label{fig:wind_parameters_pAGB}
\end{figure*}

\subsection{Wind parameters and ionising photon flux}
\label{sec:windparams}
We use the Modules for Experiments in Stellar Astrophysics ({\sc mesa}) code \citep[version r15140;][]{Paxton2011} to create stellar evolution models. The code is public\footnote{\url{https://docs.mesastar.org/en/latest/}} and collaborative, and has been updated several times throughout the years \citep[see, e.g.,][]{Paxton2013,Paxton2015,Paxton2018,Paxton2019,Jermyn2023}.

We create stellar evolution models with initial masses of 1, 2 and 3 M$_\odot$, which evolve from the ZAMS to the white dwarf (WD) phase (see Fig.~\ref{fig:HRD}). The initial abundance is set to $Z = 0.02$ and no magnetic fields or rotation are included in the calculations. The final masses of the 1, 2 and 3 M$_\odot$ stellar evolution models are 0.527, 0.593 and 0.663 M$_\odot$, respectively.

\subsubsection{The AGB phase}

The mass-loss rate during the red giant branch is modelled using the \citet{Reimers1975} formulation with standard wind efficiency of $\eta=0.5$ and the AGB mass-loss is modelled adopting the \citet{Blocker1995} scheme with a wind efficiency of $\eta=0.1$. 

The properties of the stellar wind during the evolution of the AGB wind for three models are presented in Appendix~\ref{sec:AGB}. The top panel of Fig.~\ref{fig:AGB_wind} illustrates the evolution of the mass-loss rate for the last $3\times10^{6}$\,yr of evolution during the AGB phase as obtained from {\sc mesa}. The stellar wind velocity during the AGB phase ($\varv_\mathrm{AGB}$) is calculated using the empirical relationship presented in \cite{Verbena2011}, where
\begin{equation}
    \varv_\mathrm{AGB} = 0.05 \left( \frac{L}{M}\right)^{0.57} {\mathrm{km}~\mathrm{s}^{-1}},
\label{eq:verbena}
\end{equation}
\noindent where $L$ and $M$ are the luminosity and mass of the evolving star in Solar units. The results are plotted in the bottom panel of Fig.~\ref{fig:AGB_wind} for the three stellar evolution models.

One-dimensional hydrodynamical simulations of the evolving circumstellar medium that forms when the {\sc mesa} mass-loss rates and the stellar wind velocity from Eq.~\eqref{eq:verbena} are used in the free-streaming wind prescription (Eq.~\ref{eqn:1}) were performed for the complete AGB phase. Fig.~\ref{fig:AGB_dend} illustrates the density distribution in the circumstellar gas at the end of the AGB phase, just before the onset of the post-AGB phase. As already discussed by previous authors \citep[see, e.g.,][]{Villaver2002}, most of the traces of the thermal pulses are erased due to the expansion of the circumstellar material, and a relatively simple density distribution $\sim r^{-2}$ is obtained. 

\subsubsection{The post-AGB evolution}

The post-AGB evolution of the 1, 2, and 3\,M$_\odot$ models is calculated in two ways. We first run the post-AGB evolution for H-rich stars. When the {\sc mesa} models reach $\log_{10}(T_\mathrm{eff}/\mathrm{K}) = 3.8$, heading bluewards after leaving the AGB phase, we initiate the mass-loss rate prescription given by \citet{MB2016},   
\begin{equation}
    \left( \frac{\dot{M}}{\mathrm{M}_\odot~\mathrm{yr}^{-1}} \right) = 9.778 \times 10^{-15}  \left( \frac{L}{\mathrm{L}_\odot} \right)^{1.674}  \left( \frac{Z}{Z_\odot} \right)^{2/3}.
    \label{eq:MBmlrate}
\end{equation}  
\noindent This empirical formula accounts for the influence of stellar luminosity $L$ during the post-AGB phase at a given initial metallicity, $Z$. The evolution of $\dot{M}$ with time during the first 10\,kyr of the post-AGB phase is shown in the top-left panel of Fig.~\ref{fig:wind_parameters_pAGB} (dashed lines).

\begin{figure*}
    \centering
    \includegraphics[width=0.5\linewidth]{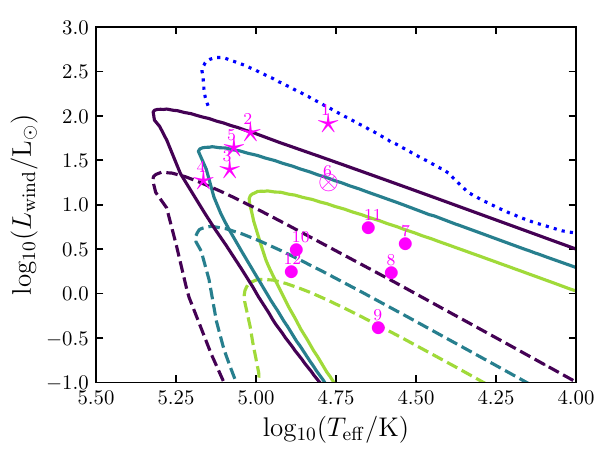}~
    \includegraphics[width=0.5\linewidth]{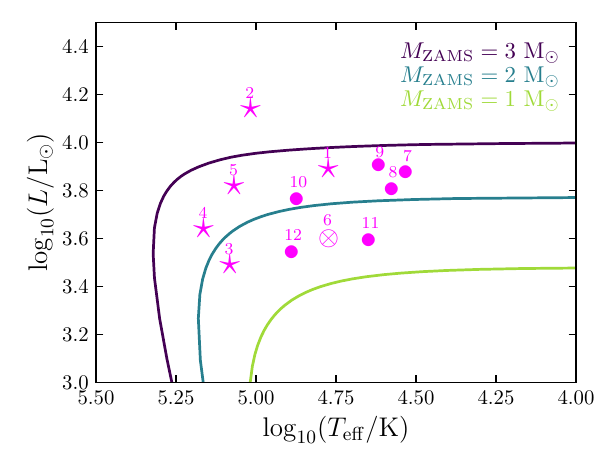}\\
    \includegraphics[width=0.5\linewidth]{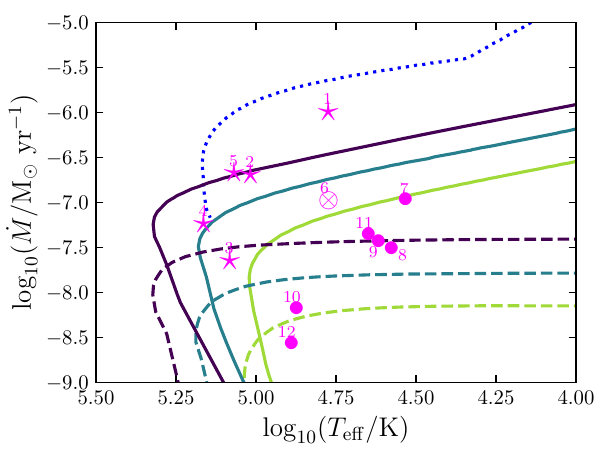}~
    \includegraphics[width=0.5\linewidth]{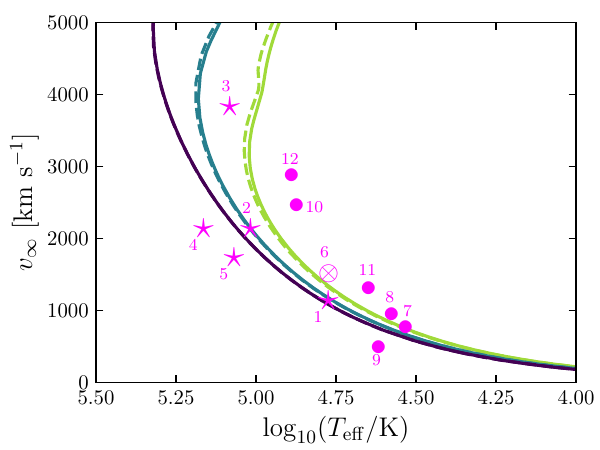}
    \caption{Evolution of the stellar wind luminosity ($L_\mathrm{wind}$ - top left), stellar luminosity ($L$ - top right), mass-loss rate ($\dot{M}$ - bottom left), and stellar wind velocity ($\varv_\infty$ - bottom right) as functions of the effective temperature $T_\mathrm{eff}$. Solid and dashed lines represent [WR] and H-rich post-AGB models, respectively. The dotted (blue) lines in the left panels represent the wind-momentum and mass-loss rate for a Post-AGB star with an increased mass-loss by a factor of 100 used in \citet{Schonberner2024}. The properties of CSPN from the literature are shown with different symbols: stars -- [WR] CSPN, bullets -- O(H) CSPN, and crossed-circle -- Of-WR (only NGC~6543). See Table~\ref{tab:XPN} for details.}
    \label{fig:Teff_obs}
\end{figure*}

A second set of post-AGB tracks is computed by adopting the mass-loss rate prescription for [WR]-type winds recently proposed by \citet{Toala2024}. These authors found a fundamental plane that correlates $\dot{M}$ with stellar luminosity $L$ and effective temperature $T_\mathrm{eff}$ according to 
\begin{equation}
\begin{aligned}
    \left(\frac{\dot{M}}{\mathrm{M}_\odot~\mathrm{yr}^{-1}}\right) = 7.76\times10^{-9} \left(\frac{L}{\mathrm{L}_\odot} \right)^{1.23}  
    \left(\frac{T_\mathrm{eff}}{\mathrm{K}} \right)^{-0.68}. 
    \label{eq:plane}
\end{aligned}   
\end{equation}
\noindent The resultant mass-loss rates for [WR]-type stars are significantly higher than for H-rich winds, which impacts the evolution of the post-AGB models. Although the post-AGB models including [WR]-type winds have the same tracks in the HR diagram as H-rich stars, their evolution is accelerated. Fig.~\ref{fig:HRD} illustrates this situation, where filled symbols correspond to times during the post-AGB evolution for [WR]-type winds, while open symbols correspond to the same times but for H-rich winds.
The top-left panel of Fig.~\ref{fig:wind_parameters_pAGB} also shows the evolution of $\dot{M}$ for [WR]-type winds (solid lines) compared to those resulting from H-rich post-AGB models.

The stellar wind terminal velocities ($\varv_\infty$) during the post-AGB phases were calculated using the analytic radiation-driven wind prescriptions of \cite{Kudritzki1989}. In particular, we applied the equations presented in Section 6 of their work, which provide explicit relations to compute the wind velocity from the stellar parameters of our evolutionary models under the assumption of a stationary, spherically symmetric, radiation-driven outflow.
The evolution of $\varv_\infty$ with time is shown in the top-right panel of Fig.~\ref{fig:wind_parameters_pAGB} and illustrates that, regardless of the mass-loss rate prescription, models with the same mass reach similar asymptotic values of $\varv_\infty$. However, the [WR]-type stars do it in shorter time scales.

\begin{figure*}
    \centering
    \includegraphics[width=0.48\linewidth]{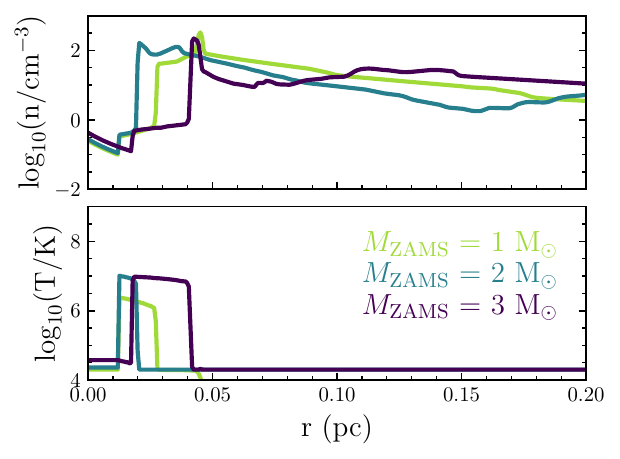}~
    \includegraphics[width=0.48\linewidth]{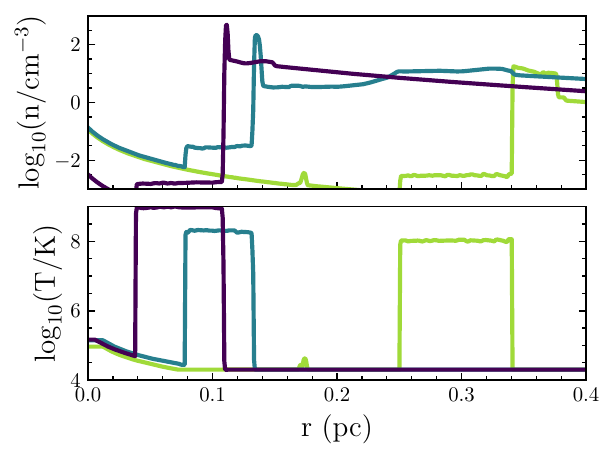}
    \caption{Radial profiles of the number density ($n$, upper panels) and plasma temperature ($T$, lower panels) for the post-AGB models 500\,yr after the beginning of the formation of the hot bubble. Left panels show the H-rich PNe evolution while the right panels correspond to the [WR] case. Different colours correspond to different ZAMS stellar masses.}
    \label{fig:density_1D}
\end{figure*}

The bottom-left panel of Fig.~\ref{fig:wind_parameters_pAGB} presents the evolution of the ionising photon rate obtained from our models. In the present paper we are mainly interested in assessing the impact of adopting [WR]-type winds on the formation of hot bubbles in PNe and, thus, a detailed estimation of the ionising photon flux, such as that presented in \citet{ToalaArthur2014}, will not be pursued in the present work. Here, we estimate the ionising photon rate $Q_0$ by integrating black body distributions for each time step in the post-AGB stellar models.

The mechanical luminosity produced by the stellar winds is
\begin{equation}
    L_\mathrm{wind} = \frac{1}{2} \dot{M} \varv_\infty^2.
\end{equation}
Its evolution with time is also presented in Fig.~\ref{fig:wind_parameters_pAGB} (bottom-right) but its relation to $T_\mathrm{eff}$ is further illustrated in the top-left panel of Fig.~\ref{fig:Teff_obs}. The mechanical wind luminosity of the H-rich post-AGB models is typical for this range of masses, where $L_\mathrm{wind}$ does not exceed 1\% of the stellar luminosity $L$ \citep[see, e.g.,][]{Perinotto2004,Steffen2008,ToalaArthur2014,ToalaArthur2016}. Given the higher $\dot{M}$ of the [WR] models, their estimated $L_\mathrm{wind}$ values are consequently higher.

Finally, to further explore the evolving relationship between the stellar and wind parameters, Fig.~\ref{fig:Teff_obs} also presents diagrams of $L$, $\dot{M}$, and $\varv_\infty$ as functions of $T_\mathrm{eff}$.

\section{Results}
\label{sec:results}

Post-AGB models with [WR]-type winds evolve more rapidly than their H-rich counterparts, primarily due to the enhanced mass-loss rates dictated by Eq.~(\ref{eq:plane}). These [WR]-type models progress faster across the Hertzsprung-Russell diagram toward higher $T_\mathrm{eff}$, with all associated parameters reflecting this accelerated evolution. Of particular importance for the formation and evolution of the hot bubble in PNe are the stellar wind terminal velocity ($\varv_{\infty}$) and the ionising photon rate ($Q_0$). The [WR]-type winds are accelerated more rapidly than their H-rich analogues, but both eventually reach comparable terminal velocities. In all cases, $\varv_\infty$ is lower for lower-mass models. Conversely, the mass-loss rate ($\dot{M}$) in [WR] models is consistently higher for lower-mass stars, a direct consequence of the negative dependence on $T_\mathrm{eff}$ in Eq.~(\ref{eq:plane}). The combination of $\dot{M}$ and $\varv_\infty$ results in higher $L_\mathrm{wind}$ values for [WR] post-AGB models compared to their H-rich counterparts.

1D simulations of the AGB and post-AGB phases can help us examine the formation and evolution of hot bubbles in PNe, and can be used to assess the impact of [WR]-type winds. There are several combinations of parameters that we are interested in. We start by mentioning that we did not cross-match models from different masses, for example, a 2\,M$_\odot$ AGB model is only taken as initial condition for the 2\,M$_\odot$ post-AGB models. 

As expected, cooling affects the properties of hot bubbles in PNe, particularly their early formation \citep[see][]{Schonberner2024}. The AGB phases and H-rich post-AGB simulations were computed adopting the H-rich cooling curve of Fig.~\ref{fig:cooling}, whilst the post-AGB models with [WR] winds use the [WR] cooling curve. In all cases, the post-AGB phase simulations start at $\log_{10}( T_\mathrm{eff}/\mathrm{K}) = 3.8$, and continue for a total of 10,000\,yr.

\subsection{Hot bubble formation}

Fig.~\ref{fig:density_1D} shows examples of the number density $n$ and gas temperature $T$ resulting from 1D simulations of the post-AGB phases for the 1, 2 and 3\,M$_\odot$ stellar evolution models. The figure shows results from H-rich (left panel) and [WR] post-AGB winds (right panel). Both figures show the presence of a low-density, high-temperature hot bubble 500 yr after the formation of the hot bubble. The $n$ and $T$ profiles are very similar to previous 1D simulations of the formation of hot bubbles of PNe without thermal conduction \citep{Stute2006,Steffen2008,ToalaArthur2014}, where a sharp contact discontinuity separates the diffuse hot bubble from the swept-up ionised nebula.

A comprehensive review of the early formation and survival of hot bubbles in PNe has recently been presented in \citet{Schonberner2024}, where the inner position of the wind shock ($R_1$) and the position of the contact discontinuity ($R_2$) evolve according to the crossing time of the free wind, the age of the hot bubble, and the radiation-cooling timescale \citep[see][]{Koo1992a, Koo1992b}. A hot bubble forms when $R_2 > R_1$. 

Given the diverse parameter space ($L_\mathrm{wind}$, $\varv_\infty$, abundances) used in our simulations, hot bubbles are not formed at the same time for all models. This can be understood using the framework developed by \citet{Koo1992a}, who analysed bubble formation in the two cases of “slow” winds and “fast” winds. These limits are separated by a critical wind velocity that comes from the requirement that the cooling time in the shocked stellar wind be comparable to the timescale for the stellar wind to sweep up its own mass. “Slow” wind bubbles are initially radiative, and consequently momentum driven, until the density behind the inner wind shock drops sufficiently such that the cooling time becomes long in comparison to the dynamical timescale. At this point the shocked stellar wind forms a hot bubble, which is energy driven. “Fast” wind bubbles are always energy driven, since the cooling time behind the wind shock is long in all cases.

\begin{figure}
    \centering
    \includegraphics[width=1.0\linewidth]{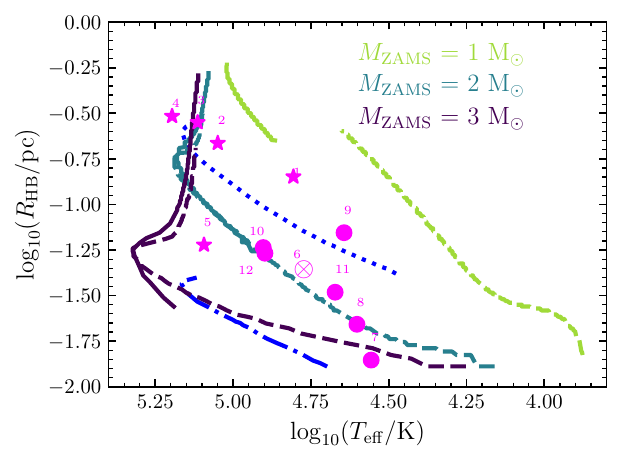}
    \caption{Evolution of hot bubble radius $R_\mathrm{HB}$ as a function of stellar effective temperature $T_\mathrm{eff}$. The solid lines show the results for post-AGB models adopting [WR] winds, while the dashed lines correspond to H-rich stars. The (blue) dotted and dash-dotted lines show the results from the 1D simulations from \citet{Schonberner2024} corresponding to their models WR100V05 and WR100V05x5.5, respectively. The (magenta) symbols represent the position of X-ray-emitting PNe listed in Table~\ref{tab:XPN}.}
    \label{fig:radius_teff}
\end{figure}

Following the same steps as equations (2.2)--(2.5) of \citet{Koo1992a}, we find that
\begin{equation}
\varv_\mathrm{crit} = 3.75\times10^5 \left( \frac{L_\mathrm{wind}}{L_\odot} n_0\right)^{1/11} \left(\Lambda_1 f(\mu) \right)^{2/11}  \ \mbox{km\,s$^{-1}$},
\label{eq:vcrit}
\end{equation}
where $\Lambda(T) = \Lambda_1 T^{-1/2}$ approximates the cooling curve for temperatures $10^5 < T < 10^6$~K \citep[see][]{Kahn1976} and $\Lambda_1 = [3.16,3160]\times 10^{-19}$~erg~cm$^3$~s$^{-1}$ for the PN and [WR] cooling curves, respectively, shown in Figure~\ref{fig:cooling}. The function $f(\mu) = \mu_\mathrm{AGB}^{1/2}/(\mu_\mathrm{ion} \mu_e \mu^{1/2}) = [0.95,0.08]$ for the H-rich and H-poor abundance sets listed in Table~\ref{tab:abund}, respectively. Finally, $n_0$ is the number density of the circumstellar medium, i.e., material expelled during the AGB, into which the bubble expands. Evaluating $\varv_\mathrm{crit}$ for the two cooling curves and corresponding abundances yields
\begin{equation}
\left(\frac{\varv_\mathrm{crit}}{\mbox{km\,s$^{-1}$}}\right) = \left\{  \begin{array}{ll}
                                       160   \left(\frac{L_\mathrm{wind}}{L_\odot} n_0\right)^{1/11}  & \mbox{PN abundances} \\
                                       360   \left(\frac{L_\mathrm{wind}}{L_\odot} n_0\right)^{1/11}  & \mbox{[WR] abundances} 
                                       \end{array}
                              \right. \ .
\label{eq:vcrit2}
\end{equation}
A hot bubble cannot start to form until $\varv_\infty > \varv_\mathrm{crit}$. Equation~(\ref{eq:vcrit2}) shows that $\varv_\mathrm{crit}$ is higher for H-poor stars not only because the abundances and cooling rate lead to a higher numerical factor but also because the wind luminosity is up to two orders of magnitude higher for a given stellar mass (see Fig.~\ref{fig:wind_parameters_pAGB}). Winds that reach their terminal velocity on a short timescale fulfil the critical velocity criteria and form hot bubbles quickly. For the 1~M$_\odot$ model, the wind velocity increases slowly for both H-rich and H-deficient cases but the H-deficient case requires the wind to reach a much higher velocity before the hot bubble begins to form.

\begin{figure}
    \centering
    \includegraphics[width=1.0\linewidth]{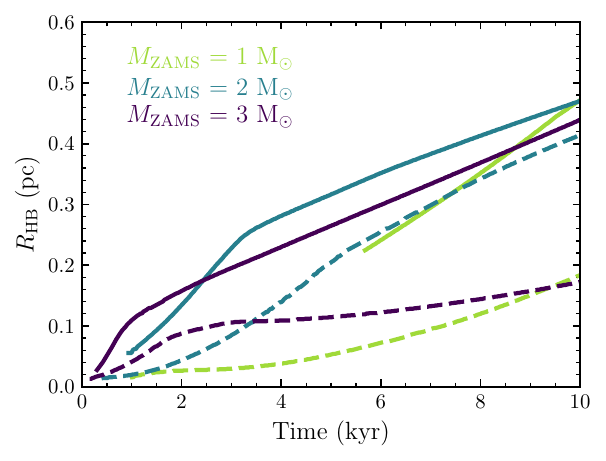}
    \caption{Evolution of the hot bubble radius $R_\mathrm{HB}$ as a function of the time during the post-AGB phase. Solid and dashed lines represent models of [WR] and H-rich post-AGB stars, respectively.}
    \label{fig:radius_age}
\end{figure}

To determine the radius of the hot bubble, we developed an algorithm that tracks the position of the inner shock throughout the simulation. We first identify strong radial gradients in density, which provide a stable numerical indicator of sharp discontinuities in the flow. However, because density jumps can also arise at contact discontinuities, the algorithm does not rely on density alone to locate the shock. Instead, each candidate discontinuity is required to satisfy an additional physical criterion: the gas temperature at that location must exceed a threshold value. This choice is motivated by the fact that only gas in this temperature regime is expected to emit soft X-rays and, moreover, that a significant increase in temperature necessarily implies a corresponding increase in pressure behind the shock ($P \propto nT$). In contrast, contact discontinuities preserve pressure continuity and therefore do not satisfy the temperature requirement. The combined density–temperature selection ensures that the detected feature corresponds to the high-pressure, inner-shock rather than to a contact discontinuity in the swept-up AGB shell. In practice, we adopt a threshold of $T \ge 5\times10^{5}\,\mathrm{K}$, although we tested alternative values ($10^{5}$ and $10^{6}\,\mathrm{K}$) and found that the resulting hot-bubble radii change only marginally. This approach allows the algorithm to robustly track the advancing inner shock throughout the evolution of the nebula.

Fig.~\ref{fig:radius_teff} shows the moment of formation and the subsequent evolution of hot bubbles in our simulations as a function of $T_\mathrm{eff}$. Hot bubbles do not always form immediately at the beginning of the post-AGB phase due to the wind velocity criterion (see Eq.~\ref{eq:vcrit2}). In particular, our simulations with [WR]-type winds and strong cooling (Fig.~\ref{fig:cooling}) form hot bubbles at a more advanced stage of the post-AGB evolution, similar to what was reported by \citet{Schonberner2024}.
For example, the hot bubble of the 3\,M$_\odot$ model with an H-rich wind forms at $\log_{10}(T_\mathrm{eff}/\mathrm{K})=4.25$, while its [WR] counterpart forms once the star reaches $\log_{10}(T_\mathrm{eff}/\mathrm{K})=5.20$.

Fig.~\ref{fig:radius_age} shows the evolution with time of the hot-bubble radius $R_\mathrm{HB}$ for all post-AGB models. The figure illustrates that the hot bubble of the 1\,M$_\odot$ [WR] model is born after about $t\sim$\,6,000\,yr of its post-AGB evolution. The 2~M$_\odot$ [WR] case forms a hot bubble after 1000\,yr. Finally, the 3~M$_\odot$ [WR] model takes only about 300 yr to generate its hot bubble.

\begin{figure}
    \includegraphics[width=1.0\linewidth]{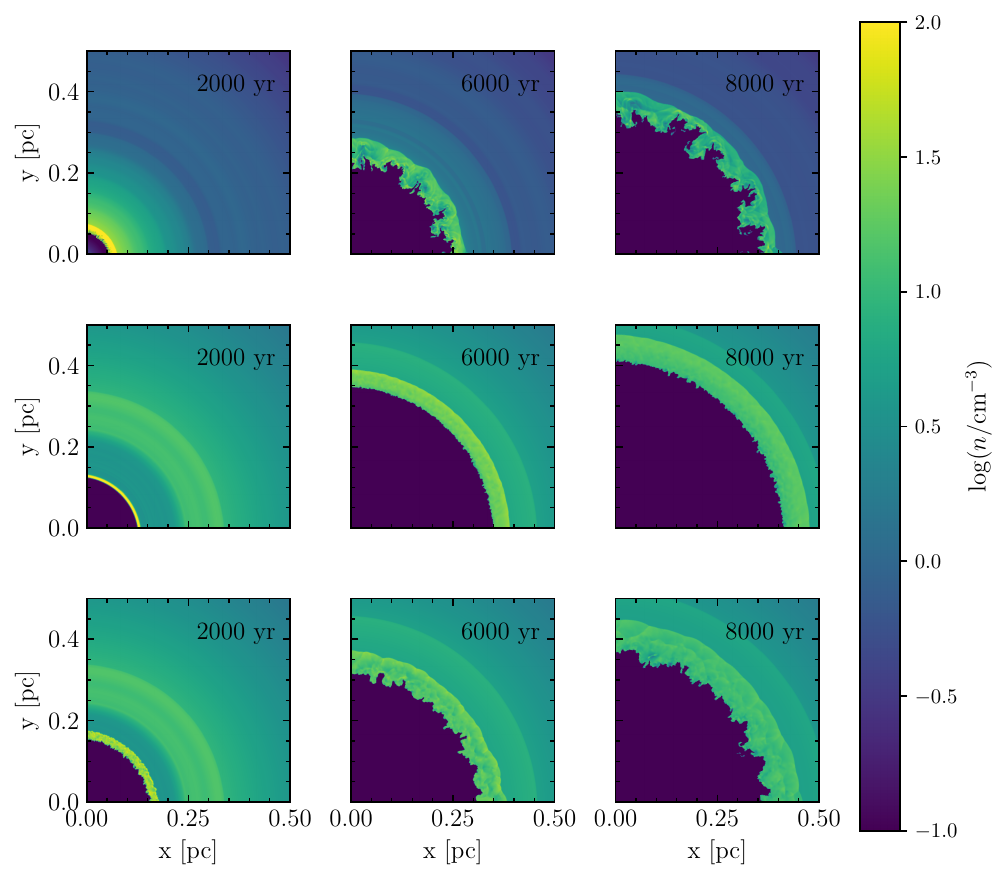}
\caption{Number density $n$ at 2000, 6000 and 8000\,yrs for the 1 (top), 2 (middle), and 3 M$_\odot$ (bottom) models with H-deficient [WR] stars.}
\label{fig:2D_WR_den}
\end{figure}

\begin{figure}
\includegraphics[width=1.0\linewidth]{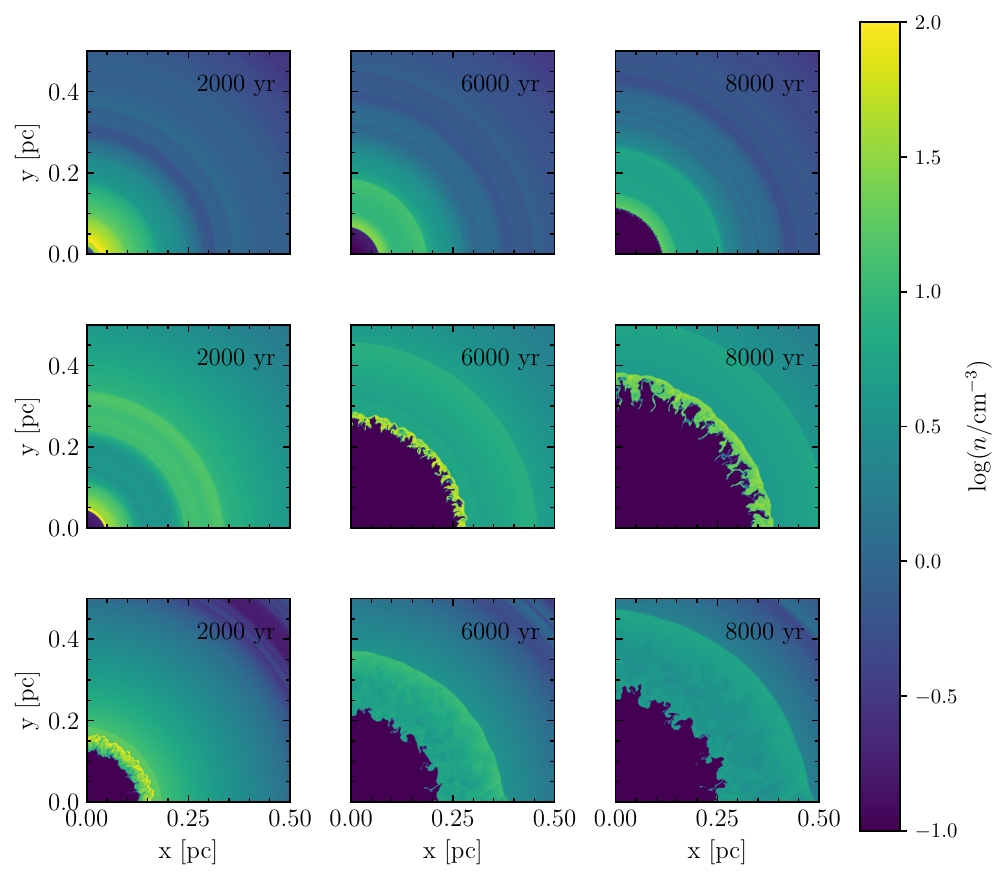}
\caption{Same as Fig.~\ref{fig:2D_WR_den} but for the H-rich post-AGB models.}
\label{fig:2D_den}
\end{figure}

\begin{figure}
    \includegraphics[width=1.0\linewidth]{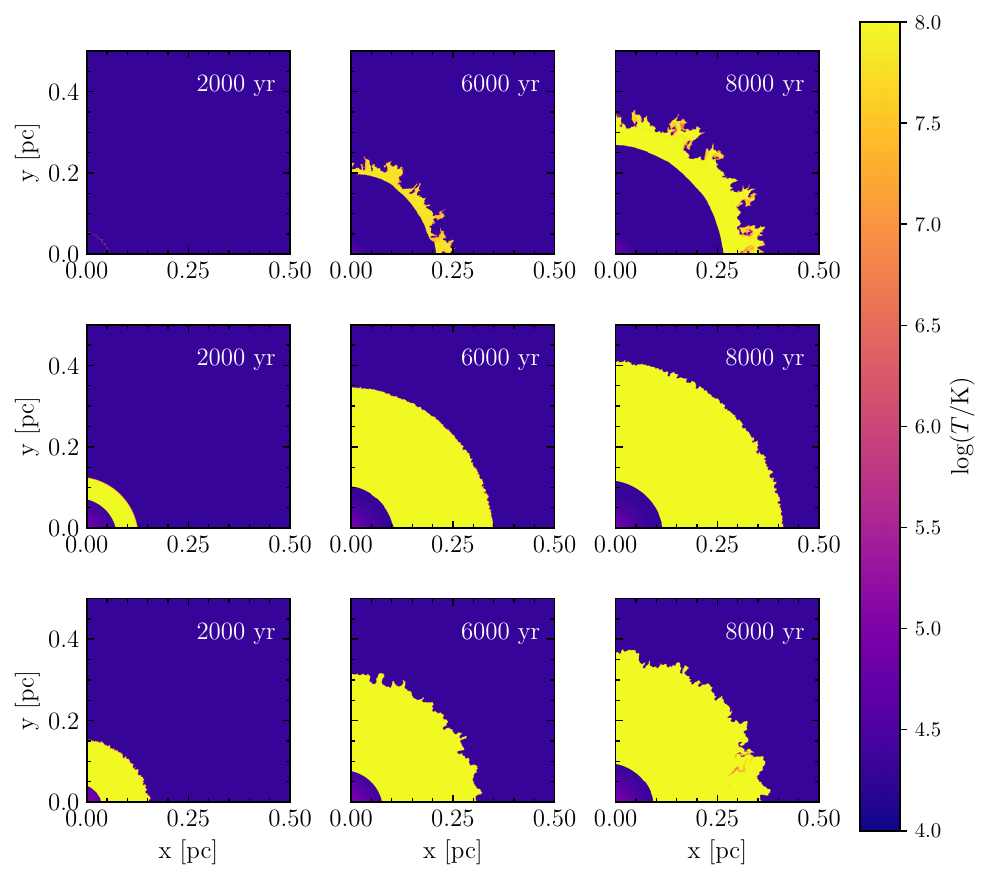}
\caption{Gas temperature $T$ at 2000, 6000 and 8000\,yrs for the 1 (top), 2 (middle), and 3 M$_\odot$ (bottom) models with H-deficient [WR] stars.}
\label{fig:2D_WR_T}
\end{figure}

\begin{figure}
    \includegraphics[width=1.0\linewidth]{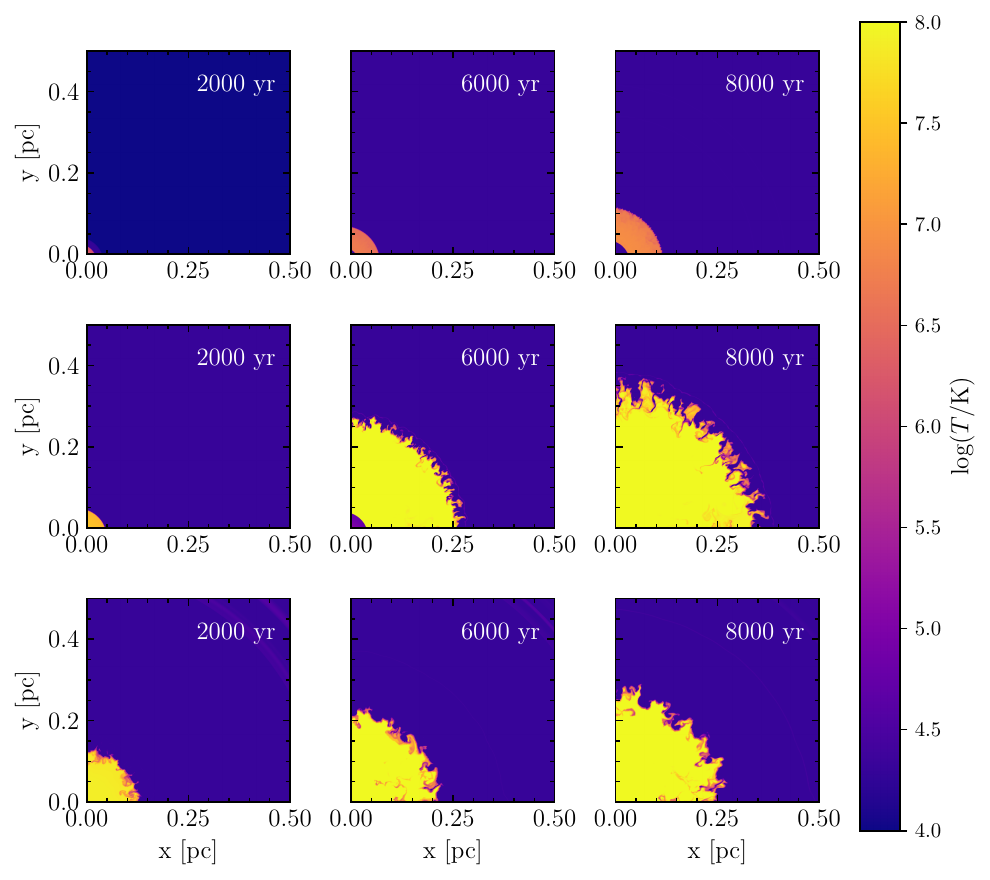}
\caption{Same as Fig.~\ref{fig:2D_WR_T} but for the H-rich post-AGB models}
\label{fig:2D_T}
\end{figure}

\begin{figure}
    \includegraphics[width=1.0\linewidth]{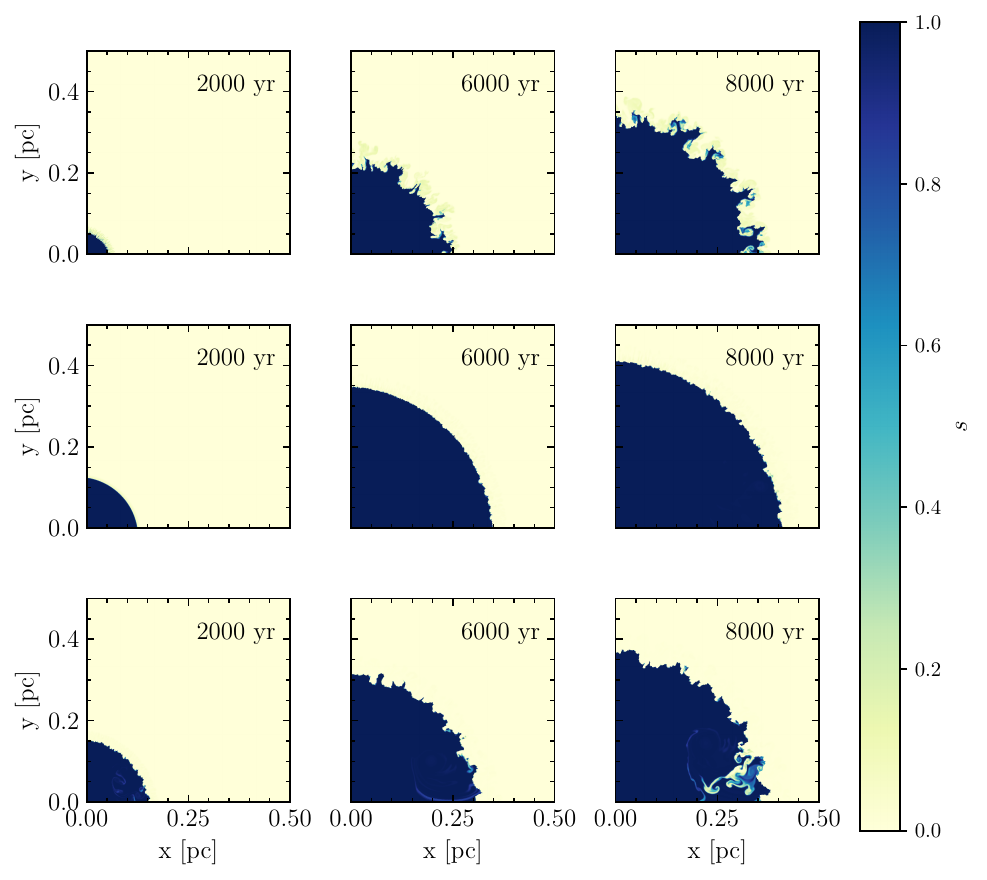}
\caption{Passive scalar $s$ at 2000, 6000 and 8000\,yrs for the 1 (top), 2 (middle), and 3 (bottom) M$_{\odot}$ with H-deficient [WR] stars.}
\label{fig:wr_scalar}
\end{figure}

\begin{figure}
    \includegraphics[width=1.0\linewidth]{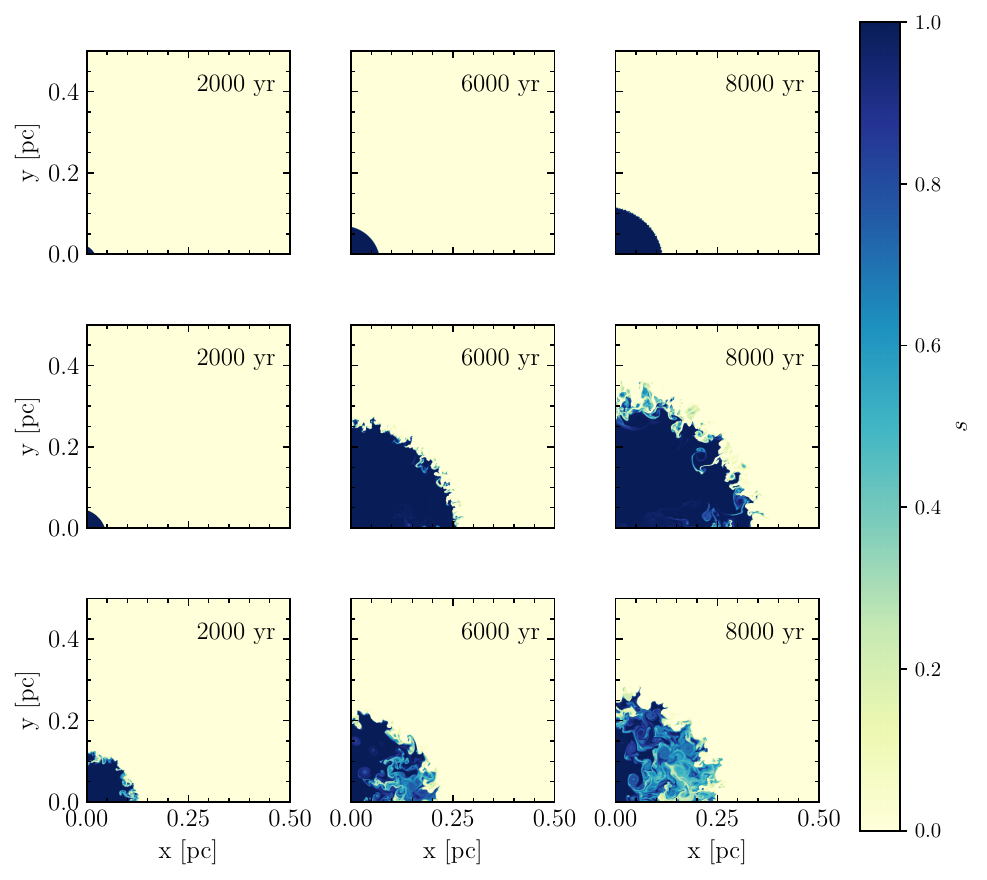}
\caption{Same as Fig.~\ref{fig:wr_scalar} but for the H-rich post-AGB models}
\label{fig:pne_scalar}
\end{figure}

The formation of hot bubbles in our 2D simulations is illustrated in Figs.~\ref{fig:2D_WR_den}--\ref{fig:2D_T}. The density structure created by H-deficient [WR] winds is presented in Fig.~\ref{fig:2D_WR_den} at 2000, 6000, and 8000\,yr after the onset of the post-AGB phase and a similar figure for the H-rich post-AGB stars is shown in Fig.~\ref{fig:2D_den}. The two figures reveal the formation of hydrodynamical instabilities in the wind-wind interaction layer. As discussed in previous works \citep{ToalaArthur2011,ToalaArthur2014}, Vishniac thin-shell instabilities \citep{Vishniac1989, Vishniac1983} develop. These are formed in radiatively cooled shells that are compressed between a pressured interior and the ram pressure of an external flow; small perturbations grow because the shell is unable to maintain lateral pressure support, leading to corrugations along its surface. Furthermore, Rayleigh-Taylor instabilities are also formed, which are developed when a dense shell is accelerated or de-accelerated by a lower-density medium, producing finger-like protrusions. The development of one type of instability over the other depends on several conditions such as the initial configuration of the AGB material, its expansion velocity, and the evolution with time of $L_\mathrm{wind}$.

The temperature structures of the 2D simulations are presented in Fig.~\ref{fig:2D_WR_T} and \ref{fig:2D_T} for the H-deficient [WR] and H-rich post-AGB models, respectively. They show that the turbulent layer of material that forms in the wind-wind interaction region has temperatures of the order $10^{6}$~K, which naturally favours soft X-ray emission \citep[see][and references therein]{ToalaArthur2018}.

We use an advected scalar parameter, $s$, to distinguish between AGB material, which is assigned the value $s=0$, and post-AGB wind material, which corresponds to $s=1$. This advected scalar enables us to follow in detail the mixing between the nebular and hot bubble material and is used to assign appropriate chemical abundances when calculating the synthetic X-ray emission. As a consequence of mixing, the turbulent region created in the wind-wind interaction zone has intermediate values \citep[see further discussion in][]{ToalaArthur2014,ToalaArthur2016}. Examples of the scalar structure of our models are presented in Fig.~\ref{fig:wr_scalar} and \ref{fig:pne_scalar}.

\subsection{Synthetic X-ray emission}
\label{sec:synthetic}

\begin{figure*}
\includegraphics[width=0.92\linewidth]{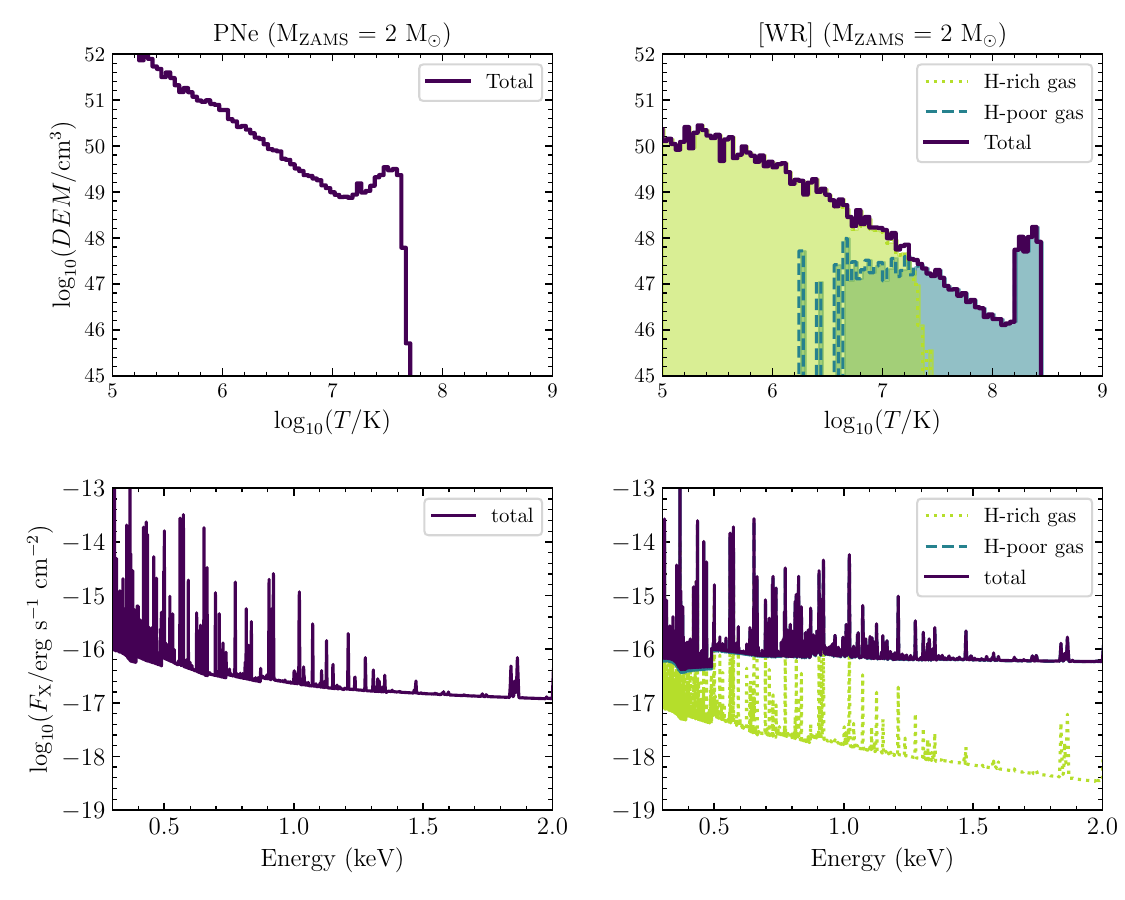}
\caption{Differential Emission Measure (DEM - top) and spectra (bottom) for the $M_\mathrm{ZAMS}=2$ M$_\odot$ models at $t = 3500$~yr. The left and right columns show the results for the H-rich and [WR] cases, respectively.
In the right-hand panels, the contributions from the H-rich and H-deficient gas are indicated with different line types.}
\label{fig:DEM_spectrum_PNe}
\end{figure*}

To elucidate the X-ray properties of the hot bubbles formed in our simulations, we followed a procedure similar to that described in \citet{ToalaArthur2016,ToalaArthur2018}. The first step is to compute the differential emission measure (DEM) from the simulations, which we define as 
\begin{equation}
    \mathrm{DEM}(T_\mathrm{b}) = \sum_{i, T_{i} \in T_\mathrm{b}} n_\mathrm{e}^2 \Delta V_{i},
\end{equation}
\noindent where $n_\mathrm{e}$ is the electron number density in cell $i$, $\Delta V_i$ is the volume of cell $i$, $T_{i}$ is the temperature of cell $i$ and the sum is performed over cells with gas temperature falling in the bin whose central temperature is $T_\mathrm{b}$. We use logarithmic binning in the temperature range $\log_{10}(T_\mathrm{b}) = 5$ to $\log_{10}(T_\mathrm{b}) = 9$ in intervals of 0.04\,dex (i.e., 100 bins).

An example of a DEM distribution obtained from the 2~M$_\odot$ simulation is presented in the top-left panel of  Fig.~\ref{fig:DEM_spectrum_PNe}, which shows the results at a time 3,500~yr after the onset of the H-rich post-AGB phase. Unsurprisingly, the DEM profiles resemble those reported in previous works of wind-blown bubbles \citep[see][]{ToalaArthur2018,Green2022,Mackey2025}.

The DEM distributions are then used in combination with the {\sc chianti} atomic database \citep[see][and references therein]{Dufresne2024} to calculate X-ray spectra. The calculations take into account the sum of all the contributions to the emission spectrum of a given temperature bin, $T_\mathrm{b}$ (e.g., collisionally excited lines, free-free, free-bound, and two-photon continua). The synthetic spectra cover the 0.3--2.0\,keV energy range, i.e., the same as that corresponding to the {\it XMM-Newton} and {\it Chandra} observations of PNe \citep[see][and references therein]{Ruiz2013}. The bottom-left panel of Fig.~\ref{fig:DEM_spectrum_PNe} shows the resultant X-ray spectrum computed from the DEM distribution displayed in the top-left panel. This spectrum is calculated using the standard H-rich PNe abundance set listed in Table~\ref{tab:abund}.

\begin{figure}
    \centering
    \includegraphics[width=.97\linewidth]{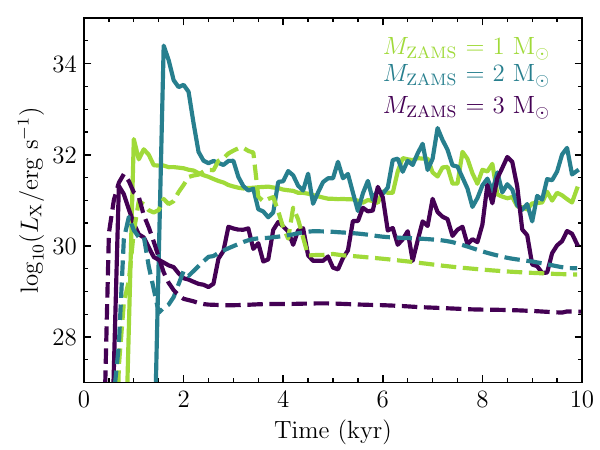}\\
    \includegraphics[width=.97\linewidth]{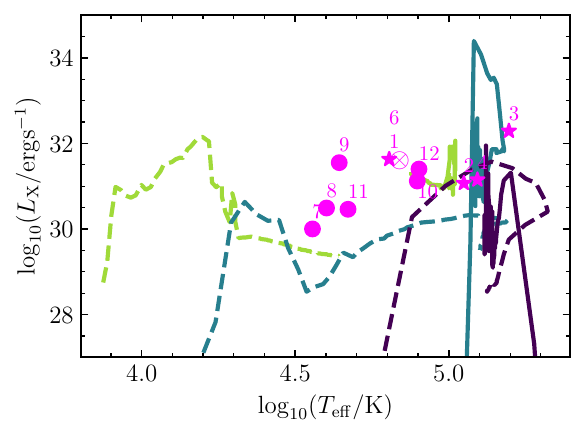}\\
    \includegraphics[width=.97\linewidth]{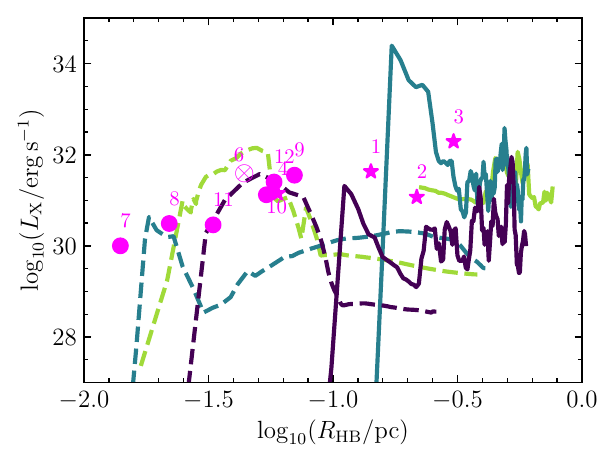}
    \caption{Luminosity of the X-ray emitting gas $L_\mathrm{X}$ as a function of time (top), effective temperature of the central star (middle), and as a function of the hot bubble radius (bottom). Solid lines correspond to models with H-deficient [WR] winds, and dashed lines are those of H-rich post-AGB stars.}
    \label{fig:x-rayluminosity}
\end{figure}

However, we note that the X-ray emissivity of hot bubbles formed by [WR]-type winds arises from a mixture of H-rich and H-poor material, and their DEM distributions must therefore be separated. We use the advected scalar parameter $s$ in our simulations to achieve this: the DEM contribution from the [WR] wind is computed using cells with $s \ge 0.5$, while the contribution from the surrounding H-rich material is derived from cells with $s < 0.5$. The top-right panel in Fig.~\ref{fig:DEM_spectrum_PNe} presents the DEM profile of a 2~M$_\odot$ simulation with a [WR] wind, which shows that the less-mixed [WR] wind exhibits a DEM distribution peaking at higher temperatures; however, its contribution to the total DEM is smaller than that of the mixed material dominated by the nebular H-rich PN gas. Synthetic spectra are computed separately, using the appropriate abundance set (see Table~\ref{tab:abund}) for each of the DEM components, which are plotted alongside the total spectrum in the bottom-right panel of Fig.~\ref{fig:DEM_spectrum_PNe}. We remark that the spectrum of the [WR] component is computed adopting the [WC] abundance set.

\begin{figure}
    \centering
    \includegraphics[width=1.0\linewidth]{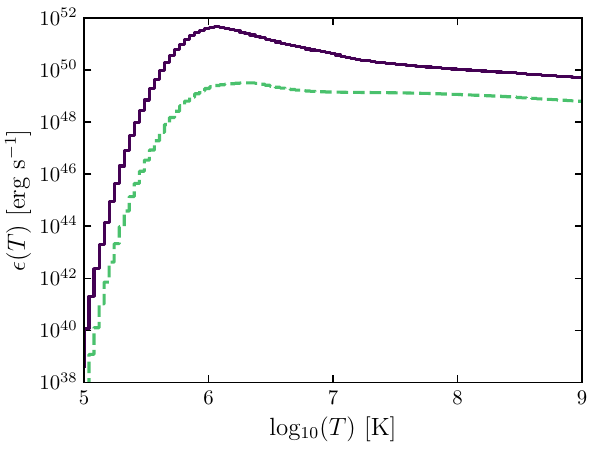}\\[-20pt]
    \includegraphics[width=1.0\linewidth]{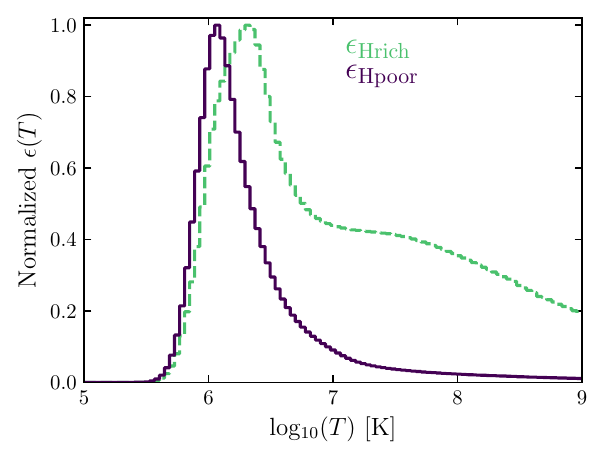}
    \caption{Emission coefficient curves for the 0.3--2.0 keV energy range obtained for standard H-rich PNe abundances ($\epsilon_\mathrm{Hrich}$, dashed line) and H-deficient [WR] abundances ($\epsilon_\mathrm{Hpoor}$, solid line).
    The bottom panel shows the same normalised curves.}
    \label{fig:temp_emiss}
\end{figure}

After computing the spectra for all simulations, we integrated them to obtain the corresponding X-ray luminosities ($L_\mathrm{X}$) in the 0.3--2.0 keV energy range. Fig.~\ref{fig:x-rayluminosity} shows the evolution of $L_\mathrm{X}$
%the X-ray luminosity 
as a function of time, $T_\mathrm{eff}$ and $R_\mathrm{HB}$ for all of our models. In all cases, the simulations produced with [WR]-type winds exhibit larger $L_\mathrm{X}$ than their H-rich counterparts. We also calculated the emissivity weighted average temperature of the X-ray emitting temperature. This was defined by \citet{ToalaArthur2016,ToalaArthur2018} to be
\begin{equation}
    T_\mathrm{X} = \frac{\int \epsilon(T) \mathrm{DEM}(T) T dT}{\int \epsilon(T) \mathrm{DEM}(T) dT}, 
    \label{eq:temp}
\end{equation}
\noindent where $\epsilon (T)$ is the emission coefficient in the X-ray band. Different abundance sets produce emissivity curves with different shapes and peak values. For the present paper, we computed two emissivity curves using {\sc chianti}, one for standard H-rich PNe abundances ($\epsilon_\mathrm{Hrich}$) and another for the H-deficient, C-rich [WR] abundance set ($\epsilon_\mathrm{Hpoor}$), where the abundances are listed in Table~\ref{tab:abund}. These are presented in Fig.~\ref{fig:temp_emiss}. 

The emissivity curve corresponding to the standard PN abundances is identical to that presented in \citet{ToalaArthur2016,ToalaArthur2018}, with a dominant peak at $\log_{10}(T/\mathrm{K}) = 6.31$. Fig.~\ref{fig:temp_emiss} corroborates that decreasing the H abundance shifts the peak of the $\epsilon(T)$ curve to lower temperatures. The emissivity curve of the H-deficient material peaks at $\log_{10}(T/\mathrm{K}) = 6.09$. The "shoulder" in the $\epsilon(T)$ curve for standard PN abundances arises from the presence of H. In contrast, for the H-deficient gas this secondary peak is suppressed due to the dominance of C at softer energies in the dominant peak. Note that the bottom panel of Fig.~\ref{fig:temp_emiss} shows the normalized version of the curves. To grasp the true difference between each $\epsilon(T)$ curve we show the true values for each curve in the top panel of Fig.~\ref{fig:temp_emiss}. As clearly seen in the figure, the H-poor emissivity is approximately 5 five orders of magnitude greater than the H-rich emissivity curve for temperatures around log$_{10}(T/\mathrm{K})\gtrsim 6$.

Consequently, in our models we need to account for the contribution from the H-rich and H-poor gas phases. For this, we define here the total emissivity-weighted X-ray temperature as:
\begin{equation}
    T_\mathrm{X} = f_\mathrm{Hrich} T_\mathrm{X,Hrich} + f_\mathrm{Hpoor} T_\mathrm{X,Hpoor},
    \label{eq:temp2}
\end{equation}
\noindent where $T_\mathrm{X,Hrich}$ and $T_\mathrm{X,Hpoor}$ have similar definitions to Eq.~(\ref{eq:temp}), that is
\begin{equation}
    T_\mathrm{X,Z} = \frac{\int \epsilon_\mathrm{Z}(T) \mathrm{DEM}_\mathrm{Z}(T) T dT}{\int \epsilon_\mathrm{Z}(T) \mathrm{DEM}_\mathrm{Z}(T) dT} 
    \label{eq:temp_Z},
\end{equation}
and the metallicity $Z$ is either "Hrich" or "Hpoor". In Eq.~(\ref{eq:temp2}), $f_\mathrm{Hrich}$ and $f_\mathrm{Hpoor}$ are the corresponding fractions of the H-poor and H-rich DEMs to the total DEM profiles. By construction, these fractions satisfy $f_\mathrm{Hrich} + f_\mathrm{Hpoor} = 1$. For example, in the case of PNe with H-rich post-AGB evolution $f_\mathrm{Hrich} = 1$ and $f_\mathrm{Hpoor} = 0$ and we recover Eq.~(\ref{eq:temp}).

\begin{figure}
    \centering
    \includegraphics[width=.93\linewidth]{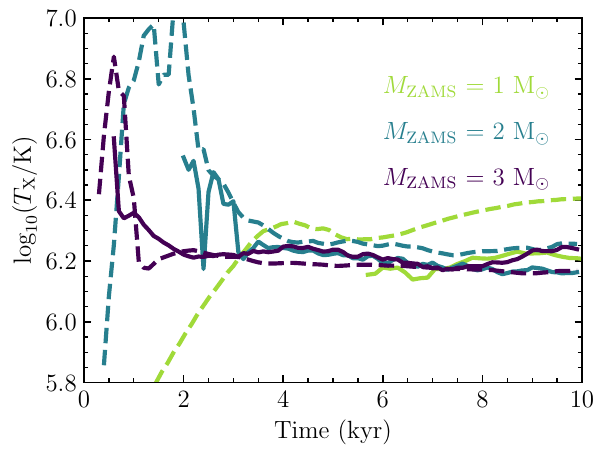}\\
    \includegraphics[width=.93\linewidth]{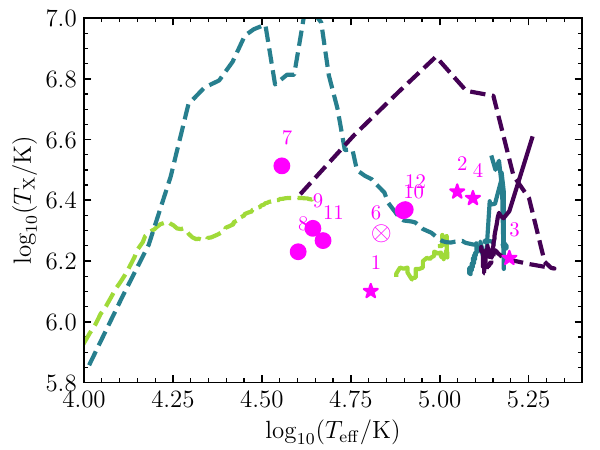}\\
    \includegraphics[width=.93\linewidth]{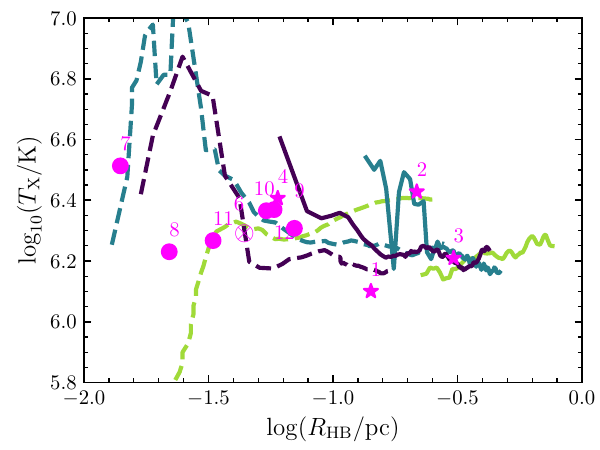}
    \caption{Emissivity weighted average plasma temperature of the X-ray-emitting gas $T_\mathrm{X}$ as a function of time (top), $T_\mathrm{eff}$ (middle), and $R_\mathrm{HB}$ (bottom). Solid lines correspond to models with H-deficient [WR] winds and dashed lines are those of H-rich post-AGB stars. Magenta symbols correspond to observed properties of PNe and their CSPN as shown in Table~\ref{tab:XPN}.}
    \label{fig:hot_bubble_temp}
\end{figure}

Following this procedure, the resultant evolution of $T_\mathrm{X}$ with time, $T_\mathrm{eff}$, and $R_\mathrm{HB}$ are presented in Fig.~\ref{fig:hot_bubble_temp} top, middle and bottom panels, respectively.

\section{Discussion}
\label{sec:discusion}

We have developed a new set of stellar evolution models with {\sc mesa} to explore how the winds from H-poor [WR] CSPN influence the formation and X-ray properties of hot bubbles within PNe. By using the stellar parameters predicted by our {\sc mesa} tracks (e.g. mass-loss rates, effective temperature, ionizing photon flux, and terminal wind velocity) as inputs for hydrodynamical simulations, we carried out a suite of 2D radiation-hydrodynamical models to assess the impact of adopting a self-consistent mass-loss prescription for [WR] stars. Our study expands upon the recent work of \citet{Schonberner2024} in three main aspects. (i) Whereas \citet{Schonberner2024} employed a single 0.595~M$_\odot$ post-AGB evolutionary track and artificially enhanced both the wind properties and the evolutionary speed to mimic the evolution of a [WR] star, our {\sc mesa} models provide stellar parameters and evolutionary timescales that arise self-consistently from stellar evolution, thereby offering a physically motivated basis for the wind energetics and temporal evolution of [WR] progenitors. (ii) Their treatment of the hot bubble structure relies on a 1D approach including thermal conduction and non-equilibrium ionisation computed through {\sc chianti}. In contrast, our simulations do not include thermal conduction but are performed in full 2D, allowing the development of turbulent mixing layers produced by non-linear hydrodynamical instabilities at the wind-wind interaction region, ultimately producing similar effects in the production of soft X-ray emission. (iii) Finally, while \citet{Schonberner2024} considered only a single stellar mass, we explore models with $M_\mathrm{ZAMS}=1$, 2, and 3~M$_\odot$ progenitors, which allows us to investigate how evolutionary speed, wind energetics (e.g. $\dot{M}$ and $\varv_\infty$), and post-AGB luminosities shape the formation and X-ray evolution of hot bubbles across a wider parameter space. This mixing-driven scenario, combined with the broader mass coverage of our models, provides a physically robust pathway to generating soft X-ray emission in hot bubbles around H-poor [WR] central stars. This complementary approach allows us to extend and further explore the implications of the results presented by \citet{Schonberner2024}, placing them within a broader and more self-consistent evolutionary framework.

\subsection{Formation of hot bubbles and stellar wind parameters}

We start by noticing that the empirical mass-loss rate relation presented by \citet{Toala2024} is strictly valid only for H-deficient WR-type winds and, in principle, our AGB models do not evolve into such configurations. The most widely accepted mechanism to form a [WR] CSPN is the born-again scenario, in which the star undergoes a late thermal pulse after leaving the AGB phase, ultimately producing an H-deficient surface composition \citep[see][and references therein]{MillerBertolami2024}. Nevertheless, adopting the wind prescription from \citet{Toala2024} allows us to assess the impact of [WR]-type winds on the formation and evolution of hot bubbles, and to compare these results with the well-established behaviour of nebulae hosting H-rich CSPN.

Despite its empirical nature, the relation proposed by \citet{Toala2024} appears to provide a reasonable approximation for the winds of [WR] CSPN. For example, 
Fig.~\ref{fig:Teff_obs} compares the stellar and wind properties of the best characterised CSPNe hosted by PNe with X-ray-emitting hot bubbles, listed in Table~\ref{tab:XPN} of Appendix~\ref{sec:XPN}. We selected PNe with H-rich CSPN with an O-type spectral classification \citep[e.g.,][]{Weidmann2020} and X-ray-emitting [WR] PNe selected from the work of \citet{Toala2024}. The latter is the most updated analysis of [WR] stars using the non-LTE stellar atmosphere {\sc PoWR} code. Fig.~\ref{fig:Teff_obs} shows that [WR] PNe systematically have larger $L_\mathrm{wind}$, consistent with our models with mass-loss rates of WR-type stars (solid lines). O-type CSPNe have smaller $L_\mathrm{wind}$ values and their hot bubbles are detected for smaller $T_\mathrm{eff}$, a result also highlighted in \citet[][]{Schonberner2024}. It is worth noting that the CSPN of NGC~6543, which has a spectral classification of Of-WR(H) exhibits a larger $L_\mathrm{wind}$ value than other O-type CSPN, more consistent with the [WR] tracks. The $\dot{M}$--$T_\mathrm{eff}$ diagram also positions [WR] CSPN above those of the O-type spectral class in accordance with the model predictions.
For comparison, the left panels of Fig.~\ref{fig:Teff_obs} show the properties of the simulations presented by \citet{Schonberner2024}, which are high above our theoretical estimations and values obtained for CSPN of X-ray-emitting PNe.

As expected, the numerical simulations presented here show that the hot bubbles carved by the impact of [WR]-type winds produce X-ray emission. The main differences between simulations evolving from H-rich post-AGB models and those resulting from [WR]-type winds are due to the stellar wind luminosities, the adopted cooling curves and the subsequent development of hydrodynamical instabilities. We confirm previous findings that enhanced cooling due to higher metallicity in the [WR] wind, combined with higher wind mechanical luminosities (see Eq.~\ref{eq:vcrit2}), inhibits the early formation of their hot bubbles \citep[see][]{Schonberner2024}. For example, the 1~M$_\odot$ [WR] model delays the formation of its hot bubble to about 6,000\,yr after the onset of the post-AGB phase. This effect helps to explain why the hot bubble radii of [WR] PNe are typically larger than those found in O-type CSPN (see Fig.~\ref{fig:radius_teff} and Table~\ref{tab:XPN}).

\subsection{X-ray properties of PN hot bubbles}

\subsubsection{Detectability of diffuse X-ray emission}

A further implication of the delayed hot-bubble formation in [WR] models is that diffuse X-ray detections should be intrinsically rare at the earliest post-AGB stages. In our simulations, the bubble only reaches the X-ray emitting regime once the CSPN has already evolved to relatively high  effective temperatures - typically $\log_{10} (T_\mathrm{eff}/\mathrm{K}) \gtrsim 4.8$ - and after the bubble has expanded beyond compact radii, about $R_\mathrm{HB}\gtrsim 0.05$\,pc. This implies that [WR] PNe with low CSPN  $T_\mathrm{eff}$ or with very compact bubbles are not expected to exhibit diffuse X-ray emission - not because the winds are weak, but because radiative cooling coupled with high wind mass-loss rates (see Eq.~\ref{eq:vcrit2}) suppresses the bubble long enough that the system has not yet entered the X-ray emitting phase. This provides a natural explanation for the absence of young, compact [WR] PNe in current X-ray samples.

\begin{figure}
    \centering
    \includegraphics[width=1.0\linewidth]{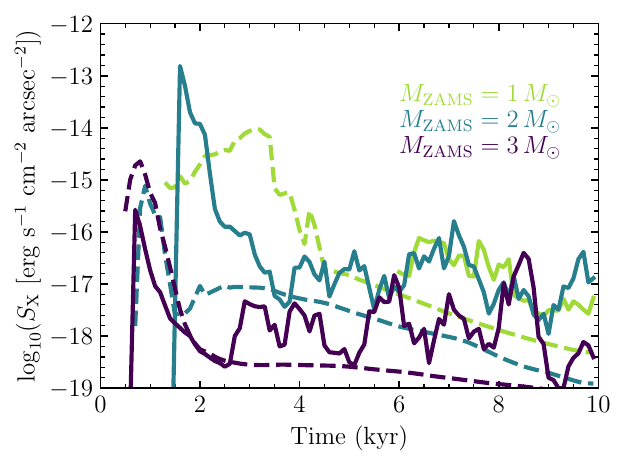}
    \includegraphics[width=1.0\linewidth]{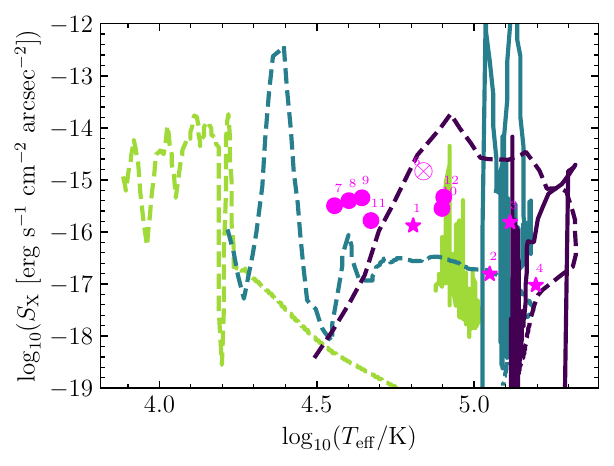}\\
    \includegraphics[width=1.0\linewidth]{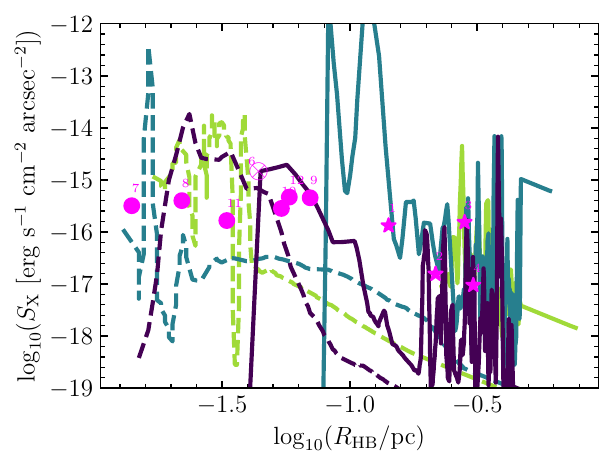}
    \caption{Evolution of the X-ray surface brightness $S_\mathrm{X}$ as a function of time (top), $T_\mathrm{eff}$ (middle), and $R_\mathrm{HB}$ (bottom). Solid lines correspond to models with H-deficient [WR] winds, and dashed lines are those of H-rich post-AGB stars.}
    \label{fig:surface_bright}
\end{figure}

Another important aspect to consider is the surface brightness of the hot bubble. Fig.~\ref{fig:surface_bright} shows the temporal evolution of the X-ray surface brightness in the 0.3–2.0\,keV energy band. After reaching a maximum, all models exhibit a rapid decline of approximately four orders of magnitude in surface brightness as the bubble expands and the gas cools. As a consequence, the window during which these objects are detectable in X-rays is limited. In particular, for the more massive models, such as the 3\,M$_\odot$ case, the phase of high surface brightness is very short and confined to times earlier than $t \lesssim 2000$\,yr. 

In the other panels of Fig.~\ref{fig:surface_bright}, we further compare the predicted X-ray surface brightness $S_\mathrm{X}$, with the observed properties of X-ray–detected PNe in the $S_\mathrm{X}$--$R_{\rm HB}$ and
$S_\mathrm{X}$--$T_{\rm eff}$ planes.
Although the model curves display a highly variable behaviour, reflecting the turbulent and rapidly evolving conditions during the initial stages of hot-bubble formation, they nevertheless suggest a general tendency for $S_\mathrm{X}$ to decrease as the bubble expands and the system evolves.

\subsubsection{Soft X-ray temperature and luminosity}

Our 2D simulations predict that both $L_\mathrm{X}$ and $T_\mathrm{X}$ reach their highest values at early times. After approximately 3,000~yr of post-AGB evolution, most simulations converge toward nearly constant values (see top panels in Figs.~\ref{fig:x-rayluminosity} and \ref{fig:hot_bubble_temp}). It is worth noting that the H-poor models tend to converge toward an almost constant value of $L_\mathrm{X}$ after the
%decreasing of its wind momentum. 
wind luminosity decreases. For instance,
Fig.~\ref{fig:wind_parameters_pAGB} shows that for the 3\,M$_{\odot}$ model, after $t \sim 1500$~yr $L_\mathrm{wind}$ decreases by an order of magnitude, which is consistent with the gradual decrease of the X-ray luminosity observed in Fig~\ref{fig:x-rayluminosity}. A similar trend is observed in the other two H-poor models.
The simulations show good agreement between the predicted values of $L_\mathrm{X}$ and $T_\mathrm{X}$ and those measured from observations of hot bubbles in PNe. The middle and bottom panels of Figs.~\ref{fig:x-rayluminosity} and \ref{fig:hot_bubble_temp} compare our synthetic predictions with X-ray observations, revealing consistent trends.

In the particular case of the 2\,M$_\odot$ models at 3500~yr, Fig.~\ref{fig:DEM_spectrum_PNe} shows that the DEM values for the temperature range $10^6$--$10^7$~K are an order of magnitude higher in the H-rich case than in the H-poor case. This leads us to expect higher X-ray luminosities from the H-rich models, however, the opposite is true (see Fig.~\ref{fig:x-rayluminosity}, top panel). This is because the emissivity coefficient for the H-deficient abundances is at least an order of magnitude higher than that for the H-rich abundances at all temperatures, as shown in the upper panel of Fig.~\ref{fig:temp_emiss}. The right-hand lower panel of Fig.~\ref{fig:DEM_spectrum_PNe} shows how the high emissivity coefficient leads to the H-deficient plasma becoming the main contributor to the X-ray flux in the 0.3--2.0~keV band, in spite of the apparently low DEM values.

\begin{figure}
    \centering
    \includegraphics[width=1.0\linewidth]{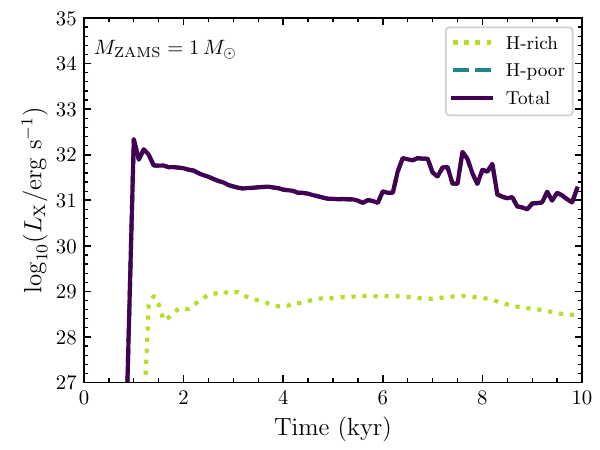}\\[-21pt]
    \includegraphics[width=1.0\linewidth]{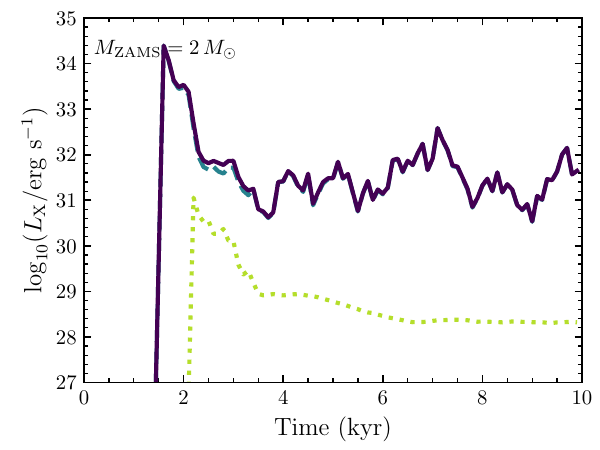}\\[-21pt]
    \includegraphics[width=1.0\linewidth]{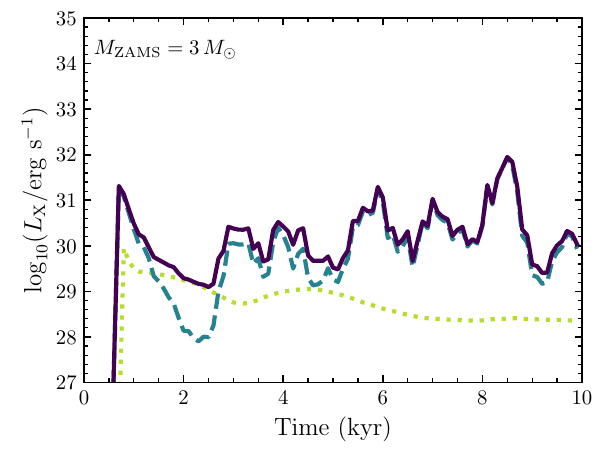}
    \caption{Evolution of the X-ray luminosity as a function of time. The top, middle and bottom panel shows the results for the 1, 2 and 3~M$_\odot$ models. Each panel shows the total X-ray luminosity (solid), the contribution from the H-poor gas (dashed), and that of the H-rich plasma (dotted).}
    \label{fig:luminosity_components}
\end{figure}

To further illustrate this behaviour, Fig.~\ref{fig:luminosity_components} presents the temporal evolution of the X-ray luminosity for the three [WR]-type models, together with the relative contributions from the H-rich and H-poor plasma components. In all cases, the total X-ray luminosity is dominated by the [WR] wind material, as $L_\mathrm{X}$ represents the sum of the individual luminosities produced by both gas components. 

On the other hand, our results show that PNe hosting [WR] central stars reach average X-ray temperatures comparable to those of PNe with H-rich winds, despite their larger wind power. Once mixing dominates the bubble evolution, the predicted plasma temperatures converge to values of $T_\mathrm{X} = [1-3] \times 10^{6}$~K, regardless of whether the system follows a [WR]-type or an H-rich post-AGB evolutionary path (see Fig.~\ref{fig:hot_bubble_temp}, top panel). This result stands in stark contrast to the predictions of \citet{Schonberner2024}, whose [WR]-type PN models systematically produce higher plasma temperatures, up to a factor of two greater than those obtained for their H-rich post-AGB counterparts. This difference arises from the multidimensional nature of our simulations. To quantify the impact of dimensionality, we computed synthetic X-ray temperatures for the corresponding 1D models and compared them directly with their 2D counterparts at matched evolutionary times. We found that 1D simulations consistently yield higher temperatures, with typical ratios of $T_{\mathrm{X,1D}}/T_{\mathrm{X,2D}} \approx 1.2$–$1.6$, and occasional deviations that can reach factors of up to $\sim$5 in our higher-mass models (2 and 3\,M$_{\odot}$). This behaviour reflects the intrinsic limitations of 1D calculations, where the absence of hydrodynamical instabilities prevents turbulent mixing between the shocked stellar wind and the swept-up nebular gas. In contrast, the multidimensional models naturally mix cooler material into the hot bubble, thereby reducing the X-ray temperature.

In 2D, the wind--wind interaction region develops hydrodynamical instabilities that promote turbulent mixing between the shocked wind and the surrounding nebular gas. This mixing efficiently transports cooler material from the swept-up shell into the hot shocked-wind region, thereby regulating its temperature even in the absence of thermal conduction. It is important to note that the behaviour of these instabilities would also change in fully 3D simulations. Several authors have reported, for example, that Rayleigh--Taylor instabilities can develop up to $\sim$35\% faster in 3D than in 2D \citep{Kane2000,Joggerst2010}. Moreover, 3D simulations show that the mixing layer develops a more complex, plume-rich, and fragmented structure than in 2D \citep{Young2001}, leading to more efficient entrainment of cold material. This suggests that the temperature of the hot bubble would likely reach its quasi-stationary state even earlier in 3D once mixing becomes dominant.

Turbulent mixing efficiently transports cooler material into the hot bubble. The entrained cold gas remains confined to the mixing interface, where it interacts with the shocked wind and drives the formation of intermediate-temperature, X-ray--emitting plasma. This behaviour is analogous to the evaporative flow that develops in a thermal conduction layer, which explains why mixing due to hydrodynamical instabilities is able to regulate the bubble temperature to the observed values even in the absence of thermal conduction. As a result, our models yield plasma temperatures that agree more closely with the observed X-ray values, whereas 1D models---unable to capture these instabilities---tend to predict systematically higher temperatures. Although 1D calculations by \citet{Schonberner2024} include non-equilibrium ionisation conditions in their X-ray calculations, their results show that NEI increases the post-shock temperature only moderately (by $\leq$ 5.6\,dex in temperature), far smaller than the differences produced by mixing. Therefore, the discrepancy between our models and 1D predictions does not arise from the use of NEI versus CIE, but rather from the absence of turbulent mixing in 1D simulations.

\begin{figure}
    \centering
    \includegraphics[width=1.0\linewidth]{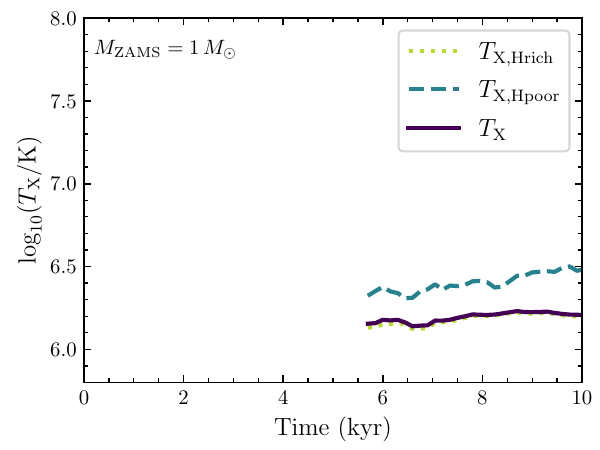}\\[-19pt]
    \includegraphics[width=1.0\linewidth]{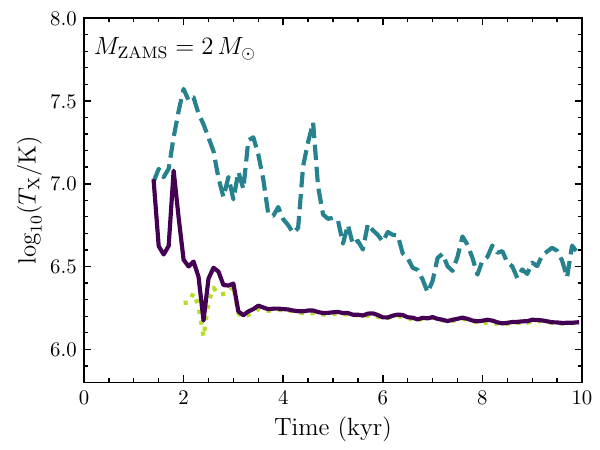}\\[-19pt]
    \includegraphics[width=1.0\linewidth]{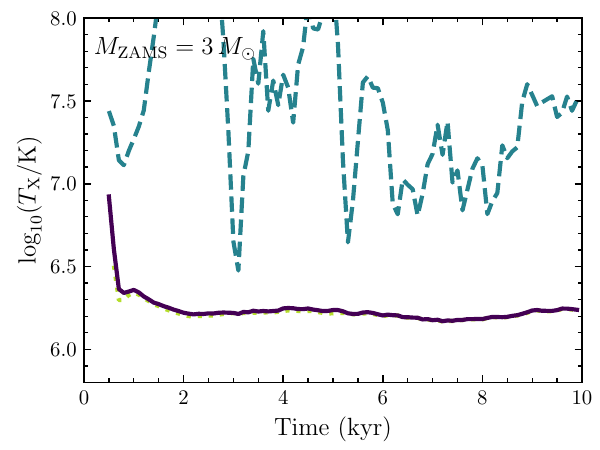}
    \caption{Evolution of the temperature of the X-ray-emitting gas in the [WR] simulations as a function of time. The top, middle and bottom panel shows the results for the 1, 2 and 3~M$_\odot$ models. Each panel shows the total X-ray emission-weighted plasma temperature ($T_\mathrm{X}$, solid), the contribution from the H-poor gas ($T_\mathrm{X,Hpoor}$, dashed), and that of the H-rich plasma ($T_\mathrm{X,Hrich}$, dotted).}
    \label{fig:temperature_components}
\end{figure}

We note that the combined X-ray temperature obtained by weighting the H-rich and H-deficient gas phases in our [WR]-type models shows the closest agreement with the single-temperature values reported by \citet{Schonberner2024}. Fig.~\ref{fig:temperature_components} illustrates the temporal evolution of $T_\mathrm{X}$, together with the individual contributions from the H-rich and H-poor gas components ($T_\mathrm{X,Hrich}$ and $T_\mathrm{X,Hpoor}$, respectively). In all cases, the plasma temperature associated with the H-deficient gas remains higher than that of the H-rich material ($T_\mathrm{X,Hpoor} > T_\mathrm{X,Hrich}$). However, the use of Eqs.~(\ref{eq:temp2}) and (\ref{eq:temp_Z}) ensures that $T_\mathrm{X}$ is properly weighted by the emissivity and the DEM profiles, following the approach described in \citet{ToalaArthur2018}. Despite being hotter, the H-poor component contributes less to the total X-ray temperature because its emissivity curve peaks outside the temperature range where most of its DEM is concentrated (see top-right panel of Fig.~\ref{fig:DEM_spectrum_PNe}). As a result, the weighting procedure gives proportionally more importance to the cooler H-rich gas originating in the mixed layer at the wind-AGB interface.

The numerical simulations presented here reinforce the view that thermal conduction is not a necessary ingredient to reproduce the observed X-ray temperatures of PNe. However, we emphasize that its inclusion would likely enhance the X-ray luminosity. Once turbulent mixing is established, the X-ray temperature no longer depends on the wind type and instead becomes regulated by the mixing physics rather than by the mechanical luminosity of the central star. In this context, two distinct evolutionary stages can be identified: at early times, when mixing has not yet developed, the plasma temperature directly reflects the shocked stellar wind; at later times, once the mixing layer is fully established, the temperature becomes insensitive to the wind power and converges toward similar values for both H-rich and [WR]-type PNe.

\section{Summary}
\label{sec:summary}

We explored the impact of [WR]-type winds in the formation, evolution, and synthetic X-ray emission from PNe. This was done through two-dimensional axially symmetric radiation-hydrodynamical simulations performed with the extensively-tested code {\sc pluto}. Our simulations followed in detail stellar evolution models from stars with initial masses of $M_\mathrm{ZAMS} = 1$, 2, and 3 M$_\odot$ created with {\sc mesa}. Two sets of post-AGB models were created, one adopted the mass-loss rate prescription tailored for the properties of H-deficient winds of [WR] stars \citep{Toala2024} and another adopting the properties of H-rich post-AGB stars \citep{MB2016}. 

Two different cooling curves were computed, one for H-deficient [WR] stars and another created adopting standard H-rich abundances of PNe. This impacts the early formation of hot bubbles in [WR] PNe as also found recently in \citet{Schonberner2024}.

We corroborate that the turbulent mixing produced by the growth of hydrodynamical instabilities at the outer edge of the hot bubble is sufficient to reproduce the observed X-ray properties. Although the precise efficiency of this mixing depends on numerical resolution and dimensionality, the instabilities themselves arise naturally from the multidimensional wind–nebula interaction. In the [WR]-type models, the combination of efficient mixing, the enhanced mechanical luminosity of the central star winds, and the strong radiative cooling naturally reproduces the X-ray properties of the observational sample.

Our main findings can be summarised as follows:

\begin{itemize}
    \item Adopting a mass-loss rate tailored for H-deficient winds of [WR] stars accelerates the evolution of low-mass stars in the HR diagram. Although the track between stars with the same $M_\mathrm{ZAMS}$ look exactly the same, the resultant stellar wind properties vary. Post-AGB models of [WR]-type stars have consistently higher mass-loss rates ($\dot{M}$) and consequently, larger mechanical luminosities ($L_\mathrm{wind}$) than the H-rich post-AGB models. The stellar wind velocity ($\varv_\infty$) of [WR]-type stars evolve to higher values faster than H-rich post-AGB models, but their final values are virtually the same.

    \item The predictions from our stellar evolution models broadly reproduce the observed properties (e.g. $L$, $T_\mathrm{eff}$, $L_\mathrm{wind}$, $\dot{M}$, and $\varv_\infty$) of a sample of observed CSPN that include those of [WR] and O-type spectral class.

    \item We corroborate previous findings that hot bubbles created by the impact of [WR]-type stars are delayed in comparison with their counterparts from H-rich post-AGB winds \citep[see][]{Schonberner2024}. This is purely a consequence of radiative cooling by H-deficient, C-rich gas being about 3 orders of magnitude larger as previously estimated \citep[e.g.,][]{Mellema2002}. This effect can delay the hot bubble formation up to 6,000 yr of evolution during the post-AGB phase of the 1~M$_\odot$ [WR] model, but it is less dramatic in higher mass models.

    \item In all cases, hydrodynamical instabilities at the outer edge of the hot bubble dominate the mixing of material between the ionised nebular gas and the shocked stellar wind. The details of clump and filament formation vary from model to model and depend on (i) the circumstellar density structure left by the AGB phase, (ii) the evolution of the post-AGB wind, and (iii) the adopted cooling curve. Although the small-scale morphology of the mixing layer is resolution dependent and not fully spatially converged, the onset of the instabilities and the global mixing efficiency are robust features of multidimensional simulations. Moreover, the integrated X-ray properties ($L_{\mathrm{X}}$ and $T_{\mathrm{X}}$) converge much more rapidly than the fine structure of the flow and are therefore not strongly affected by increasing the spatial resolution. Consequently, each model naturally develops different mixing patterns and distinct X-ray evolutionary histories.

    \item At early times, the estimated X-ray-emitting plasma temperature ($T_\mathrm{X}$) and luminosity ($L_\mathrm{X}$) reflect the turbulent mixing effects at the edge of the hot bubble. However, once mixing has established, both quantities reach basal values. 

    \item The winds from [WR] models have a major impact in the estimated luminosity in the 0.3--2.0 keV energy range. Models with [WR] winds have higher $L_\mathrm{X}$ than their H-rich post-AGB counterparts.
    
    \item In contrast to previous works, all of our simulations predict $T_\mathrm{X}=[1-3]\times10^{6}$~K regardless of whether the system follows a [WR]-type or an H-rich post-AGB evolutionary path. Our models are consistent with estimated $T_\mathrm{X}$ from X-ray-emitting PNe, including [WR] PNe. Although the wind power of [WR] post-AGB models are considerably higher, cooling is equally important.

\end{itemize}

We successfully reproduce the observed properties of X-ray-emitting hot bubbles in PNe, including those hosting [WR]-type CSPN. This result corroborates our previous findings that hydrodynamical instabilities at the wind-wind interaction region generate sufficient mixing to account for the observed soft X-ray emission in all PNe. Although we cannot rule out the possible contribution of thermal conduction, our results reinforce the conclusion that soft X-ray emission can arise even in the presence of magnetic fields, without the need for additional transport mechanisms.

Finally, we remark that the empirical mass-loss rate relation proposed by \citet{Toala2024} is formally valid only for H-deficient, WR-type winds, whereas the post-AGB models used here do not naturally evolve into such configurations. To further investigate the X-ray properties of PNe hosting [WR] stars, an important next step is to predict their X-ray emission using hydrodynamical simulations that follow a star through a born-again event. Nevertheless, the wind approximation adopted here provides a robust and practical way for modelling the winds of [WR] CSPNe.

\section*{Acknowledgements} 

The authors thank the referee for taking the time reviewing our original manuscript. R.O.D.~thanks UNAM DGAPA (Mexico) for a postdoc fellowship. R.O.D., J.A.T.~and J.B.R.G.~acknowledge support from the UNAM PAPIIT project IN102324. R.K.~acknowledges financial support via the Heisenberg Research Grant funded by the Deutsche Forschungsgemeinschaft (DFG, German Research Foundation) under grant no.~KU 2849/9, project no.~445783058.
This work has made extensive use of NASA's Astrophysics Data System.


\section*{DATA AVAILABILITY}

The data resulted from our theoretical calculations will be shared on reasonable request to the corresponding author.

\appendix

\section{AGB wind evolution}
\label{sec:AGB}

\begin{figure}
    \centering
    \includegraphics[width=1.0\linewidth]{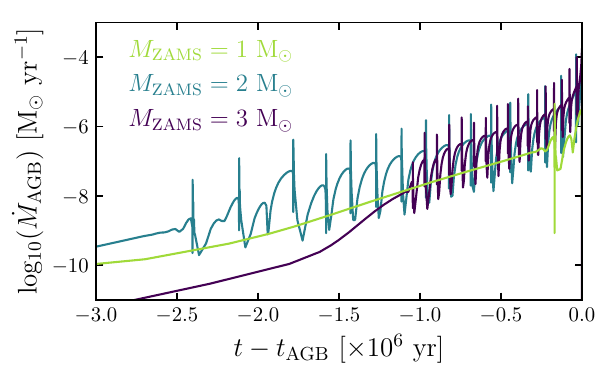}\\[-22pt]
    \includegraphics[width=1.0\linewidth]{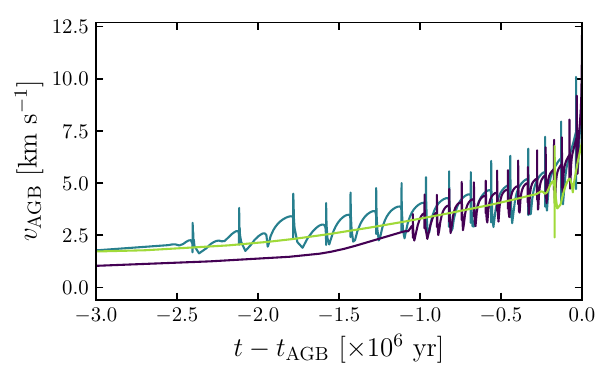}\\
    \caption{Evolution with time of the mass-loss rate ($\dot{M}_\mathrm{AGB}$, top) and stellar wind velocity ($\varv_\mathrm{AGB}$, bottom) during the last 3$\times10^{6}$ yr of evolution of the AGB phase before the onset of the post-AGB phase. The panels show results for the 1, 2, and 3 M$_\odot$ models.}
    \label{fig:AGB_wind}
\end{figure}

\begin{figure}
    \centering
    \includegraphics[width=1.0\linewidth]{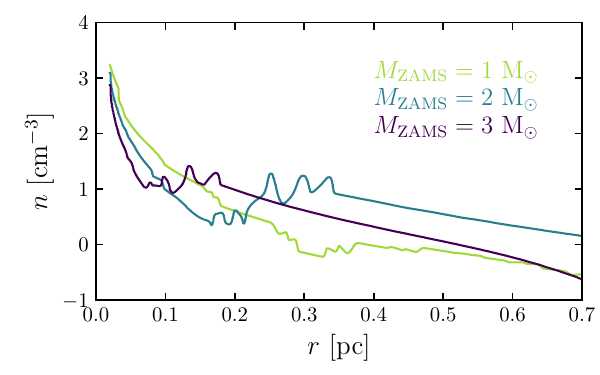}
    \caption{Distribution of the number density $n$ as a function of radius $r$ at the end of the AGB phase for the 1, 2, and 3 M$_\odot$ models. Only the inner 0.7 pc of the simulations are shown.}
    \label{fig:AGB_dend}
\end{figure}

Fig.~\ref{fig:AGB_wind} presents the final $3\times10^{6}$ yr of evolution with time of the mass-loss rate ($\dot{M}_\mathrm{AGB}$) and stellar wind velocity ($\varv_\mathrm{AGB}$) of the AGB phase. $\dot{M}_\mathrm{AGB}$ is directly obtained from the stellar evolution model, with $\varv_\mathrm{AGB}$ computed using the empirical relationship presented in \citet{Verbena2011} (see Eq.~\ref{eq:verbena}). The figure depicts the so-called thermally-pulsating AGB phase of evolution of the 2 and 3 M$_\odot$ stellar evolution models.

These parameters are used in 1D simulations to create the initial conditions of the post-AGB models. The final configuration of the total number density $n$ at the end of the AGB phase for the 2 and 3 M$_\odot$ models is illustrated in Fig.~\ref{fig:AGB_dend}. We only show the inner 0.7 pc of the simulation, which is the size of the initial box of the 2D simulations. These simulations do not include an ionising photon rate. 

Both density profiles in Fig.~\ref{fig:AGB_dend} exhibit the recent interaction of the material ejected from the last thermal pulses. At least 5 peaks are shown in the 2 M$_\odot$ profile at distances of $\approx0.2-0.3$ pc, whilst the profile of the 3 M$_\odot$ model shows 4 peaks at distances of $\approx$0.1--0.2 pc.

\section{Stellar, wind parameters, and X-ray properties of PNe}
\label{sec:XPN}

We have gathered all the stellar wind properties of CSPN with diffuse X-ray emission in Table~\ref{tab:XPN}. The table lists the properties of [WR] and O-type CSPN, in addition to the O-WR star of the Cat's Eye Nebula (NGC\,6543). We adopted the distances ($d$) reported by \citet{BJ2021}, which were computed by the analysis of {\it Gaia} data. Note that we only included PNe with [WR] stars that have been recently modelled by the updated version of the {\sc PoWR} code in \citet{Toala2024}. The properties of other CSPNe were retrieved from the literature but have been rescaled to the distances adopted here. All references are also listed in Table~\ref{tab:XPN}.

\begin{table*}
\begin{center}
\caption{Properties of hot bubbles in PNe and their CSPN from the literature. All distances were taken from \citet{BJ2021}. The spectral type, stellar and wind properties of [WR]-type CSPN ($T_\mathrm{eff}$, $L$, $\dot{M}$, and $\varv_\infty$) are taken from the most recent stellar atmosphere analysis presented in \citet{Toala2024}.}
\setlength{\tabcolsep}{\tabcolsep}  
\renewcommand{\arraystretch}{1}
\begin{tabular}{clccccccccccl}
\hline
Label & Object  & Sp. Type & $d$   & $T_\mathrm{eff}$ & $\log_{10}(L)$         & $\log_{10}(\dot{M})$   & $\varv_\infty$  & $\log_{10}(L_\mathrm{wind})$ & $\log_{10}(L_\mathrm{X})$ & $T_\mathrm{X}$ & $R_\mathrm{HB}$ & Ref. \\
&         &          & (kpc) & (kK)             & (L$_\odot$) & (M$_\odot$ yr$^{-1}$) & (km s$^{-1}$) & (L$_\odot$) & (L$_\odot$) & ($10^{6}$ K) & (pc) & \\
\hline
%[S71d]   & [WO3]    & \\
1  & NGC\,40  & [WC8]    & 1.63 & 64  & 3.85 & $-$6.10 & 1000 & ~~~1.80 & $-1.95\pm0.44$ & $1.26\pm0.26$ & 0.142 & (1,2) \\
2  & NGC~1501 & [WO4]    & 1.66 & 112 & 4.10 & $-$6.80 & 2000 & ~~~1.70 & $-2.51\pm0.45$ & $2.68\pm0.91$ & 0.217 & (1,2) \\
3  & NGC~2371 & [WO1]    & 1.66 & 130 & 3.45 & $-$7.75 & 3700 & ~~~1.28 & \dots          & \dots         & 0.282 & (1)   \\
4  & NGC~5189 & [WO1]    & 1.40 & 157 & 3.60 & $-$7.34 & 2000 & ~~~1.16 & $-1.29\pm0.10$ & $1.62\pm0.23$ & 0.305 & (1,3) \\
5  & NGC~6369 & [WO3]    & 1.03 & 124 & 3.78 & $-$6.78 & 1600 & ~~~1.52 & $-2.43\pm0.77$ & $2.55\pm1.24$ & 0.060 & (3) \\
\hline
6  & NGC~6543 & Of-WR    & 1.37 & 64  & 3.57 & $-$7.05 & 1420 & ~~~1.15 & $-2.14\pm0.05$ & $1.85\pm0.06$ & 0.040 & (2,4) \\
\hline
7  & IC~418   & O(H)f    & 1.32 & 36  & 3.84 & $-$7.06 & 650  & ~~~0.47 & $-3.58\pm0.13$ & $3.26\pm0.77$ & 0.014 & (2,5) \\
8  & IC~4593  & O(H)5f   & 2.28 & 40  & 3.77 & $-$7.60 & 830  & ~~~0.14 & $-3.09\pm0.19$ & $1.70$        & 0.022 & (2,6) \\
9  & NGC~2392 & O(H)6f   & 1.74 & 44  & 3.87 & $-$7.52 & 370  & $-$0.49 & $-2.03\pm0.08$ & $2.03\pm0.10$ & 0.070 & (2,4,7) \\
10 & NGC~3242 & O(H)     & 1.28 & 79  & 3.73 & $-$8.27 & 2350 & ~~~0.37 & $-2.46\pm0.05$ & $2.32\pm0.13$ & 0.054 & (2,8) \\
11& NGC~6826 &  O(H)3f  & 1.29 & 47  & 3.56 & $-$7.44 & 1200 & ~~~0.62 & $-3.12\pm0.21$ & $1.85\pm0.39$ & 0.033 & (2,8) \\
12 & NGC~7009 & O(H)     & 1.20 & 82  & 3.51 & $-$8.65 & 2770 & ~~~0.13 & $-2.18\pm0.05$ & $2.34\pm0.12$ & 0.058 & (2,9) \\
\hline
\end{tabular}
\label{tab:XPN}
\begin{flushleft}
References: (1) \citet{Toala2024}, (2) R. Montez Jr. et al. (in prep.), (3) \citet{Toala2019}, (4) \citet{HeraldBianchi2011}, (5) \citet{Morisset2009}, (6) \citet{Toala2020}, (7) \citet{Guerrero2019}, (8) \citet{Pauldrach2004}, (9) H.~Todt (priv. comm.)
\end{flushleft}
\end{center}
\end{table*}

\end{document}